\documentclass[reqno]{amsart}  % , draft
\usepackage{amssymb}
\usepackage{upref}
\usepackage{hyperref}
\newcommand{\bbI}{{\mathbb{I}}}
\newcommand{\bbN}{{\mathbb{N}}}
\newcommand{\bbR}{{\mathbb{R}}}

\newcommand{\bbZ}{{\mathbb{Z}}}
\newcommand{\bbC}{{\mathbb{C}}}

\newcommand{\calB}{{\mathcal B}}

\newcommand{\calF}{{\mathcal F}}
\newcommand{\calG}{{\mathcal G}}
\newcommand{\calH}{{\mathcal H}}

\newcommand{\calM}{{\mathcal M}}

\renewcommand{\gg}{p}
\newcommand{\dott}{\,\cdot\,}
\newcommand{\hatt}{\widehat}  % use \hat in subscripts and upperlimits of int.

\newcommand{\no}{\nonumber}
\newcommand{\lb}{\label}
\newcommand{\f}{\frac}

\newcommand{\ol}{\overline}

\newcommand{\Oh}{O}

\newcommand{\spec}{\text{\rm{spec}}}

\newcommand{\dom}{\text{\rm{dom}}}
\newcommand{\bi}{\bibitem}

\newcommand{\eps}{\varepsilon}

\DeclareMathOperator{\tr}{tr}
\DeclareMathOperator{\tl}{Tl}
\DeclareMathOperator{\sTl}{s-Tl}

\DeclareMathOperator{\stl}{s-Tl}
\DeclareMathOperator{\htl}{\widehat{Tl}}
\DeclareMathOperator{\TL}{Tl}

\newcommand{\funksj}{\ell^\infty(\bbZ)}

\allowdisplaybreaks
\numberwithin{equation}{section}
\newtheorem{theorem}{Theorem}[section]
\newtheorem{lemma}[theorem]{Lemma}

\newtheorem{hypothesis}[theorem]{Hypothesis}
\newtheorem{remark}[theorem]{Remark}
\theoremstyle{definition}

\newcommand{\abs}[1]{\lvert#1\rvert}

\begin{document}
\title[Conservation Laws and Hamiltonians for  the Toda Hierarchy]{Local Conservation Laws and the Hamiltonian Formalism for the Toda Hierarchy Revisited}
% Information for  author
\author[F.\ Gesztesy]{Fritz Gesztesy}
\address{Department of Mathematics,
University of Missouri,
Columbia, MO 65211, USA}
\email{fritz@math.missouri.edu}
\urladdr{http://www.math.missouri.edu/people/fgesztesy.html}
%\thanks{}
% Information for  author
\author[H.\ Holden]{Helge Holden}
\address{Department of Mathematical Sciences,
Norwegian University of
Science and Technology, NO--7491 Trondheim, Norway}
\email{holden@math.ntnu.no}
\urladdr{www.math.ntnu.no/\~{}holden/}
\thanks{Research supported in part by the Research Council of Norway,
the US National Science Foundation under Grant No.\ DMS-0405526.}
%\thanks{}
%----- end authors
%\dedicatory{}
\date{August 15, 2006}
%\date{\today}
\subjclass[2000]{Primary 37K10, 37K20, 47B36; Secondary 35Q58, 37K60.}
\keywords{Toda hierarchy, Hamiltonian formalism, conservation laws.}
%\thanks{To appear in {\it }}
%%%%%%%%%%%%%%%%%%%%%%%%%%%%%%%%%%%% 
\begin{abstract}
We revisit an elementary recursive approach to local conservation laws and the Hamiltonian formalism for the Toda hierarchy. 
\end{abstract}
%%%%%%%%%%%%%%%%%%%%%%%%%%%%%%%%%%%%
\maketitle

%%%%%%%%%%%%%%%%%%%%%%%%%%%%%%%%%%%%
%%%%%%%%%%%%%%%%%%%%%%%%%%%%%%%%%%%%
\section{Introduction}\lb{s1}
%%%%%%%%%%%%%%%%%%%%%%%%%%%%%%%%%%%%
%%%%%%%%%%%%%%%%%%%%%%%%%%%%%%%%%%%%

In this paper we revisit local conservation laws and the Hamiltonian formalism for the Toda hierarchy.  Consider sequences 
$\{a(n,t_0), b(n,t_0)\}_{n\in\bbZ}\in\ell^\infty (\bbZ)$ satisfying some decay assumptions as $n\to\pm\infty$ to be specified later, parametrized by the deformation (time) parameter $t_0\in \bbR$, that are solutions of the Toda equations
\begin{equation}
\begin{pmatrix}a_{t_0} -a(b^+ -b)\\ b_{t_0}
-2\big(a^2-(a^-)^2\big)\end{pmatrix} =0.
\end{equation}
Here $c^{\pm}$ denote shifts, that is, $c^{\pm}(n)=c(n\pm 1)$, $n\in\bbZ$.  Then clearly
\begin{equation}
\partial_{t_0}\sum_{n\in\bbZ} \ln (a(t_0,n)) = \partial_{t_0}\sum_{n\in\bbZ} b(t_0,n)=0,
\end{equation}
provided one has sufficient decay of the sequences $a,b$. Additional calculations also show that 
\begin{equation}
\partial_{t_0}\sum_{n\in\bbZ} \bigg(a(t_0,n)^2+ \frac12 b(t_0,n)^2\bigg)=0.
\end{equation}
Indeed, one can show the existence of a sequence $\{\rho_{+,j}\}_{j\in\bbN}$ of polynomials of $a,b$ and certain shifts thereof, with the property that the lattice sum is time-independent,  
\begin{equation}
\partial_{t_0}\sum_{n\in\bbZ} \rho_{+,j}(t_0,n)=0, \quad j\in\bbN.
\end{equation}
This result is obtained by deriving local conservation laws of the type
\begin{equation}
\partial_{t_0} \rho_{+,j}+(S^+ -I)J_{+,0,j}=0, \quad j\in\bbN,
\end{equation} 
for certain polynomials $J_{+,0,j}$  of $a,b$ and certain shifts thereof. The polynomials 
$J_{+,0,j}$ are constructed via a straightforward and explicit recursion relation. 

The above analysis extends to the full Toda hierarchy as follows. The $p$th equation, $p\in\bbN_0=\bbN\cup\{0\}$, in the Toda hierarchy is given by
\begin{equation}
\tl_p(a,b) = \begin{pmatrix}  a_{t_p}-a(f_{p+1}^+-f_{p+1}) \\[1mm]
b_{t_p}+g_{p+1}-g_{p+1}^- \end{pmatrix}=0,
\quad  t_p\in \bbR,  % \lb{3.2.13}
\end{equation}
where $f_\ell, g_\ell$ are carefully designed polynomial expressions of $a,b$ and certain shifts thereof. Recursively, they are given by
\begin{align}
& f_0=1, \quad g_0=-c_1, %\lb{1.2.4a} 
\\
& 2f_{\ell+1}+g_\ell+g_\ell^{-}-2bf_\ell=0, \quad \ell\in \bbN_0, %\lb{1.2.4b} 
\\
&
g_{\ell+1}-g_{\ell+1}^{-}+2\big(a^2f_{\ell}^{+}-(a^{-})^2f_\ell^{-}\big)-b(g_\ell-g_\ell^{-})=0,
\quad \ell \in \bbN_{0}, %\lb{1.2.4c}
\end{align}
where $c_1\in\bbC$ is a constant. On each level in the recursion an arbitrary constant 
$c_\ell \in\bbC$ is introduced. In the homogenous case, where all these constants 
$c_\ell$, $\ell\in\bbN$, are set equal to zero, a hat is added in the notation, that is,  
$\hat f_\ell, \hat g_\ell$, etc., denote the corresponding homogeneous quantities.  The homogeneous coefficients $\hat f_\ell, \hat g_\ell$ can also be  expressed explicitly in terms of appropriate matrix elements of powers of the Lax difference expression $L$ 
defined in \eqref{h1.2.2} below (cf.\ Lemma \ref{l3.4.3}).  The conserved densities 
$\rho_{+,j}$ are independent of the equation in the hierarchy while the currents $J_{+,p,j}$ depend on $p$; thus one finds (cf.\ Theorem \ref{thm3.1})
\begin{equation}
\partial_{t_p} \rho_{+,j}+(S^+ -I)J_{+,p,j}=0, \quad j\in\bbN, \; p\in\bbN_0.
\end{equation}
Our approach is based on a careful analysis of the high-energy expansion of the Green's function for the operator $H$ in $\ell^2(\bbZ)$ corresponding to the difference  expression
\begin{equation}
L=aS^{+}+a^{-}S^{-}+b. \lb{h1.2.2}
\end{equation}
In particular, we do not depend on real-valued sequences $a,b$, and thus the operator  $H$ is permitted to be non-self-adjoint.

In addition, we study the Hamiltonian formalism for the Toda hierachy. Proving the relations (cf.\ Lemma \ref{lemma5.1})
\begin{align}
\frac{\delta \hat{f}_\ell}{\delta a}&=-\frac{\ell}{a}\hat{g}_{\ell-1}, \quad \ell\in\bbN,  
%\lb{h5.16}
\\
\frac{\delta \hat{f}_\ell}{\delta b}&= \ell \hat{f}_{\ell-1}, \quad \ell\in\bbN,  % \lb{h5.15} 
\end{align}
one infers that the $p$th equation in the Tode hierarchy can be written as (cf.\ Theorem \ref{thm:5.4})
\begin{equation}
\tl_p(a,b)= \begin{pmatrix}a_{t_p}\\  b_{t_p}\end{pmatrix}
 - D \begin{pmatrix}(\nabla I_p)_a\\[1mm]  (\nabla I_p)_b\end{pmatrix} =0, \quad 
 p\in\bbN_{-1}   \lb{h1.1.14}
\end{equation}
(here $\bbN_{-1}=\bbN\cup\{0,-1\}$), where the functionals $I_{p}$ are given by
\begin{equation}
I_{-1}=\widehat I_{-1}, \quad I_p=\sum_{k=0}^p c_{p-k} \widehat I_k, \; p\in\bbN_0, \quad 
\widehat I_{q}= \frac{1}{q+2}\sum_{n\in\bbZ} \big(\hat{f}_{q+2}(n)-\lambda_{q+2}\big), 
\; q\in\bbN_{-1}. 
%\lb{h5.24a}
\end{equation}
Here $D$ is an explicit $2\times 2$ matrix-valued difference expression (cf.\ 
\eqref{h5.2.1}) and  $\lambda_{p+2}$ are counterterms necessary to guarantee convergence of the infinite series involved. Furthermore, we show that any $I_{r}$ is conserved by the Hamiltonian flows in 
\eqref{h1.1.14}(cf.\  Theorem \ref{thm5.5}), that is,
\begin{equation}
\frac{d I_r}{dt_p} =0, \quad p,r\in\bbN_{-1}.  % \lb{h9.60}
\end{equation} 
Moreover, for  general sequences $a,b$ ( i.e., not assuming that they satisfy an equation in the Toda hierarchy),  we show in Theorem \ref{ht5.6} that
\begin{equation}
\{\hatt I_p, \hatt I_r\}=0,  \quad  p, r \in \bbN_{-1}, %\lb{h9.61}
\end{equation}
for a suitably defined Poisson bracket $\{\dott,\dott\}$ (see \eqref{h5.8}).

Finally, we also show the existence of a second Hamiltonian structure for the Toda  hierarchy (cf.\ \eqref{h5.52},  \eqref{h5.53}), and sketch the possible extension of this formalism to the case of almost periodic sequences $a, b$ on $\bbZ$.

The Toda hierarchy and its  interpretation as a Hamiltonian system has of course received enormous attention in the past. Without being able to provide a comprehensive review of the vast literature we mention that Lie algebraic approaches to Hamiltonian and gradient structures in the asymmetric nonperiodic Toda flows, Poisson maps, etc., are studied, for instance, in  \cite{BG98}. Master symmetries for the finite nonperiodic TL are treated in \cite{Da93}, the 2nd Hamiltonian structure for the periodic Toda lattice is discussed in \cite{Be97}, and multiple Hamiltonian structures for various Bogoyavlensky--Toda systems, Poisson manifolds, master symmetries, recursion operators, etc., can be found in \cite{Da00} and \cite{DF02}. Moreover, Poisson brackets are studied in 
 \cite[Sect.\ III.4]{FT87}, \cite{FM76}, \cite[Sect.\ II.4]{NMPZ84},and 
constants of motion appeared for instance, in \cite[Sect.\ 7.3]{Ei83}, 
\cite[Sect.\ III.2.4]{FT87}, \cite{Fl74}, \cite{GH94}, \cite[Sect.\ 13.4]{Te00}, and 
\cite[Sect.\ 3.7]{To89}.  Fairly recently, an elementary approach to the infinite sequence of Toda hierarchy conservation laws was presented in \cite{ZC02} (but it seems to lack an explicit and recursive structure). Moreover, a Hamiltonian formalism for the Toda hierarchy, based on recursion operators and hence on an algebraic formalism familiar from the one involving formal pseudodifferential expressions  in connection with the 
Gelfand--Dickey hierarchy, was recently developed in \cite{ZC02a}.

Still, we emphasize that our recursive and most elementary approach to local conservation laws of the infinite Toda hierarchy does not appear to exist in the literature. Moreover, our treatment of Poisson brackets and variational derivatives, and their connections with the diagonal Green's function of the underlying Lax operator, now puts the Toda  hierarchy on precisely the same level as the KdV hierarchy with respect to this particular aspect of the Hamiltonian formalism.    

%%%%%%%%%%%%%%%%%%%%%%%%%%%%%%%%%%%%
%%%%%%%%%%%%%%%%%%%%%%%%%%%%%%%%%%%%
\section{The Toda hierarchy in a nutshell}
\label{s2}
%%%%%%%%%%%%%%%%%%%%%%%%%%%%%%%%%%%%
%%%%%%%%%%%%%%%%%%%%%%%%%%%%%%%%%%%%

In this section we briefly review the recursive construction of
the Toda hierarchy and associated Lax pairs following \cite{BGHT98}, \cite[Sect.\ 1.2]{GH07}, and
\cite[Ch.\ 12]{Te00}.

Throughout this section we make the following assumption:
%%%%%%%%%%%%%%%%%%%%%%%%%%%%%%%%%%%%% 
\begin{hypothesis} \lb{h2.1}
Suppose
\begin{equation}
a, \, b \in \ell^\infty (\bbZ), \quad a(n)\neq 0 \, \text{ for all $n\in\bbZ$}.   \lb{1.2.1}
\end{equation}
\end{hypothesis}
%%%%%%%%%%%%%%%%%%%%%%%%%%%%%%%%%%%%% 
We consider the second-order Jacobi difference expression
\begin{equation}
L=aS^{+}+a^{-}S^{-}+b, \lb{1.2.2}
\end{equation}
where $S^{\pm}$ denote the shift operators
\begin{equation}
(S^{\pm}f)(n)=f^{\pm}(n)=f(n{\pm}1), \quad n\in\bbZ, \; f \in
\bbC^\bbZ.    \lb{1.2.3}
\end{equation} 
Here $\bbC^J$ denotes the set of complex-valued sequences indexed by 
$J\subseteq\bbZ$.

To construct the stationary Toda hierarchy we need a second
difference expression of order $2\gg +2, \, \gg \in {\bbN_{0}},$
defined recursively in the following. We take the quickest route
to the construction of $P_{2\gg +2}$, and hence to the Toda  
hierarchy, by starting from the recursion relations
\eqref{1.2.4a}--\eqref{1.2.4c} below.

Define $\{{f_\ell}\}_{\ell \in \bbN_0}$ and $\{{g_\ell}\}_{\ell \in \bbN_0}$
recursively by
\begin{align}
& f_0=1, \quad g_0=-c_1, \lb{1.2.4a} \\
& 2f_{\ell+1}+g_\ell+g_\ell^{-}-2bf_\ell=0, \quad \ell\in \bbN_0, \lb{1.2.4b} \\
&
g_{\ell+1}-g_{\ell+1}^{-}+2\big(a^2f_{\ell}^{+}-(a^{-})^2f_\ell^{-}\big)-b(g_\ell-g_\ell^{-})=0,
\quad \ell \in \bbN_{0}. \lb{1.2.4c}
\end{align}
Explicitly, one finds
\begin{align}
& f_0=1, \no \\
& f_1=b+c_1, \no \\
& f_2=a^2+(a^{-})^2+b^2+c_1b+c_2, \, \text{ etc.,} \lb{1.2.5}\\
& g_0=-c_1, \no\\
& g_1=-2a^2-c_2, \no \\
& g_2=-2a^2(b^+ +b)+c_1(-2 a^2)-c_3, \, \text{ etc.} \no
\end{align}
Here $\{ c_j\}_{j \in \bbN}$ denote undetermined summation
constants which naturally arise when solving
\eqref{1.2.4a}--\eqref{1.2.4c}.

Subsequently, it will be convenient to also introduce the
corresponding homogeneous coefficients $\hat f_\ell$ and $\hat g_\ell$,
defined by vanishing of the constants $c_k,\, k \in \bbN$,
\begin{align}
& \hat f_0=1, \quad \hat f_\ell=f_\ell \big|_{c_k=0, \, k=1,\dots,\ell},
\quad \ell\in\bbN, \no \\
& \hat g_\ell=g_\ell \big|_{c_k=0, \, k=1,\dots,\ell+1}, \quad \ell\in\bbN_0.
\lb{1.2.6}
\end{align}
Hence,
\begin{equation}
f_\ell=\sum_{k=0}^{\ell}c_{\ell-k}\hat{f}_{k},\quad
g_\ell=\sum_{k=1}^{\ell}c_{\ell-k}\hat{g}_{k}-c_{\ell+1},\quad \ell\in\bbN_0,
\lb{1.2.7}
\end{equation}
introducing
\begin{equation}
c_0=1. \lb{1.2.7aa}
\end{equation}

The following explicit characterization of $\hat{f}_\ell$, $\hat{g}_\ell$ turns out to be  useful.
%%%%%%%%%%%%%%%%%%%%%% %%%%%%%%%%%%%%%%% 
\begin{lemma}\lb{l3.4.3}
Suppose $a,b$ satisfy Hypothesis \ref{h2.1}. Then 
the homogeneous coefficients $\{\hat{f}_\ell\}_{\ell\in\bbN_{0}}$ 
and $\{\hat{g}_\ell\}_{\ell\in\bbN_{0}}$
satisfy
\begin{align}
\hat{f}_\ell(n) & = (\delta_n, L^\ell \delta_n)_{\ell^2(\bbZ)}, \quad \ell\in\bbN_{0}, 
 \lb{3.4.37}\\
\hat{g}_\ell(n) & = -2 a(n) (\delta_{n+1}, L^\ell \delta_n)_{\ell^2(\bbZ)}, \quad
\ell\in\bbN_{0},  \lb{3.4.38}
\end{align}
where $\delta_n =\{\delta_{n,m}\}_{m\in\bbZ}$, $n\in\bbZ$, and 
$(\dott,\dott)_{\ell^2(\bbZ)}$ denotes the usual scalar product in $\ell^2(\bbZ)$. 
\end{lemma}
%%%%%%%%%%%%%%
For the proof see \cite{BGHT98}.

Next we define difference expressions $P_{2\gg+2}$ of order
$2\gg+2$ by
\begin{equation}
P_{2\gg+2}=-L^{\gg+1}+\sum_{\ell=0}^{\gg} \Big(
g_\ell+2af_\ell S^{+}\Big)L^{\gg-\ell}+f_{\gg+1},\quad \gg\in\bbN_0.
\lb{1.2.8}
\end{equation}
Introducing the corresponding homogeneous difference expressions
$\hatt{P}_{2p+2}$ defined by
\begin{equation}
\hatt{P}_{2\ell+2}=P_{2\ell+2}\big|_{c_k=0,\,k=1,\dots,\ell}, \quad
\ell\in\bbN_0, \lb{3.2.9A}
\end{equation}
one finds
\begin{equation}
P_{2p+2} =\sum_{\ell=0}^p c_{p-\ell} \hatt{P}_{2\ell+2}. \lb{3.2.29}
\end{equation}
Using the recursion relations \eqref{1.2.4a}--\eqref{1.2.4c}, the
commutator of $P_{2\gg+2}$ and $L$ can be explicitly computed and
one obtains
\begin{align}
[P_{2\gg+2},L]=&-a\big(g_\gg^{+}+g_\gg+f_{\gg+1}^{+}+f_{\gg+1}-2b^{+}f_\gg^{+}
\big)S^{+} \no \\
&+2\big(-b(g_{\gg}+f_{\gg+1})+a^2f_\gg^{+}
-(a^{-})^2f_{\gg}^{-}+b^2f_\gg\big) \no \\
&-a^{-}\big(g_\gg+g_\gg^{-}+f_{\gg+1}+f_{\gg+1}^{-}-2bf_\gg
\big)S^{-}, \quad \gg\in\bbN_0. \lb{1.2.8a}
\end{align}
In particular, $(L,P_{2\gg+2})$ represents the celebrated
\textit{Lax pair} of the Toda hierarchy. 
%------- remark
\begin{remark}
Varying $\gg\in\bbN_0$,
the stationary Toda hierarchy is then defined in terms of the
vanishing of the commutator of $P_{2\gg+2}$ and $L$ in
\eqref{1.2.8a} by
\begin{equation}
[P_{2\gg+2}, L] =\sTl_\gg(a,b)=0,\quad \gg\in\bbN_0. \lb{1.2.8aa}
\end{equation}
Equation \eqref{1.2.8aa} can be shown to be equivalent to
\begin{equation}
\stl_p(a,b)= \begin{pmatrix} f_{p+1}^+-f_{p+1} \\
g_{p+1}-g_{p+1}^- \end{pmatrix}=0, \quad p\in\bbN_0. \lb{3.2.29d}
\end{equation}
Specifically, one obtains
\begin{align}
\stl_0 (a,b) &=  \begin{pmatrix} b^+ -b\\
2\big((a^-)^2 -a^2\big)\end{pmatrix} =0, \no  \\%[3mm]
\stl_1 (a,b) &=  \begin{pmatrix} (a^+)^2 -(a^-)^2
+(b^+)^2 -b^2 \\ 2(a^-)^2 (b+b^-) -2a^2 (b^+ +b)
\end{pmatrix}  \lb{3.2.16a} \\
& \quad \, + c_1 \begin{pmatrix} b^+ -b \\ 2 \big((a^-)^2 -a^2\big)
\end{pmatrix} =0, \text{ etc.,} \no
\end{align}
for the first few equations of the stationary Toda hierarchy. 
\end{remark}
%--------- end remark

Next, we introduce polynomials $F_\gg(z,n)$ and $G_{\gg+1}(z,n)$
of degree $\gg$ and $\gg+1$ with respect to the spectral parameter
$z\in\bbC$ by
\begin{align}
F_\gg(z,n)&=\sum_{\ell=0}^\gg f_{\gg-\ell}(n)z^\ell , \lb{1.2.11a} \\
G_{\gg+1}(z,n)&=-z^{\gg+1}+\sum_{\ell=0}^\gg
g_{\gg-\ell}(n)z^\ell +f_{\gg+1}(n). \lb{1.2.11b}
\end{align}
Explicitly, one obtains
\begin{align}
F_0&=1, \no \\
F_1&=z+b+c_1, \no \\
F_2&=z^2+bz+a^2+(a^-)^2+b^2+c_1(z+b)+c_2, \, \text{ etc.,} \lb{1.2.12}
\\ G_1&=-z+b, \no \\
G_2&=-z^2+(a^-)^2-a^2+b^2+c_1(-z+b), \, \text{ etc.} \no
\end{align}

Next, we study the restriction of the difference expression
$P_{2\gg+2}$ to the two-dimensional  kernel (i.e., the formal null
space in an algebraic sense as opposed to the functional analytic
one) of $(L-z)$. More precisely, let
\begin{align}
& \text{ker}(L-z)=\{\psi \colon \bbZ \to \bbC\cup\{\infty\} \mid
(L-z)\psi=0\}. \lb{1.2.13a}
\end{align}
Then \eqref{1.2.8} implies
\begin{equation}
P_{2\gg+2}\mid_{\text{ker}(L-z)}=\big(2aF_\gg(z)S^{+}+G_{\gg+1}(z)\big)\big|_{\text{ker}(L-z)}.
\lb{1.2.13}
\end{equation}

Next we turn to the time-dependent Toda hierarchy. For that
purpose the functions $a$ and $b$ are now considered as functions of both
the lattice point and time. For each equation in the  hierarchy, that is,
for  each $p$, we introduce a deformation (time) parameter
$t_p\in\bbR$ in $a, b$, replacing $a(\dott), b(\dott)$ by $a(\dott,t_{p}), b(\dott,t_p)$.
The second-order  difference expression $L$ (cf.\ \eqref{1.2.2}) now reads
\begin{equation}
L(t_p) = a(\dott,t_p) S^+ +a^-(\dott,t_p) S^- +b(\dott,t_p). \lb{3.2.11A}
\end{equation}
The quantities $\{f_\ell\}_{\ell\in\bbN_0}$,
$\{g_\ell\}_{\ell\in\bbN_0}$, and $P_{2p+2}$, $p\in\bbN_0$ are still
defined by \eqref{1.2.4a}--\eqref{1.2.4c} and \eqref{1.2.8},
respectively. The time-dependent Toda hierarchy is then obtained by
imposing the Lax commutator equations
\begin{equation}
L_{t_p}(t_p) -[P_{2p+2}(t_p), L(t_p)]=0, \quad t_p\in\bbR, \lb{3.2.11}
\end{equation}
varying $p\in\bbN_0$. Relation \eqref{3.2.11} implies
\begin{align}
&\left(a_{t_p} +a (g_p^+
+g_p + f_{p+1}^+ + f_{p+1} - 2b^+ f_p^+ )\right) S^+ \no\\
& \quad -\left(-b_{t_p} +2 \big(-b (g_p + f_{p+1}) +a^2 f_p^+
-(a^-)^2 f^-_p +b^2 f_p\big)\right) \lb{3.2.11aaa}\\
& \quad +\left(a_{t_p} +a (g_p^+
+g_p + f_{p+1}^+ + f_{p+1} - 2b^+ f_p^+ )\right)^- S^-=0.\no
\end{align}
In addition one computes
\begin{align}
0&=L_{t_p} -[P_{2p+2}, L] \no \\
&=\left(a_{t_p}-a(f_{p+1}^+-f_{p+1})\right)S^+
-\left(-b_{t_p}-g_{p+1}+g_{p+1}^- \right) \no \\
& \quad +\left(a_{t_p}-a(f_{p+1}^+-f_{p+1})\right)^-S^-.
\lb{3.2.11a} %\\
%&=\tl_p (a,b)_1 S^+ -\tl_p (a,b)_2+\tl_p
%(a,b)^-_1 S^-, \quad p\in\bbN_{0}, \lb{3.2.11a}
\end{align}
Varying $p\in\bbN_0$, the collection of evolution equations
\begin{equation}
\tl_p(a,b) = \begin{pmatrix}  a_{t_p}-a(f_{p+1}^+-f_{p+1}) \\[1mm]
b_{t_p}+g_{p+1}-g_{p+1}^- \end{pmatrix}=0,
\quad (n,t_p)\in\bbZ\times\bbR, \; p\in\bbN_0 \lb{3.2.13}
\end{equation}
then defines the time-dependent Toda hierarchy. Explicitly,
\begin{align}
\tl_0 (a,b) = & \begin{pmatrix}a_{t_0} -a(b^+ -b)\\ b_{t_0}
-2\big(a^2-(a^-)^2\big)\end{pmatrix} =0, \no  \\
\tl_1 (a,b) = & \begin{pmatrix}a_{t_1} -a\big((a^+)^2 -(a^-)^2
+(b^+)^2 -b^2\big)\\b_{t_1} +2(a^-)^2 (b+b^-) -2a^2 (b^+ +b) \end{pmatrix}
\lb{3.2.16}  \\
&+ c_1 \begin{pmatrix} -a(b^+ -b)\\ -2\big(a^2-(a^-)^2\big)\end{pmatrix}
=0, \, \text{ etc.,} \no
\end{align}
represent the first few equations of the time-dependent Toda hierarchy.
The system of equations, $\tl_0(a,b)=0$, is of course \textit{the} Toda
system.

The corresponding homogeneous Toda
equations obtained by taking all summation constants equal to zero,
$c_\ell=0$, $\ell=1,\dots, p$, are then denoted by
\begin{equation}
\htl_p (a,b)= \tl_p(a,b) \big|_{c_\ell =0,\,
\ell=1,\dots, p}. \lb{3.2.18}
\end{equation}

Restricting the Lax relation \eqref{3.2.11} to the
kernel $\ker(L-z)$ one finds that
\begin{align}
0 & = \big( L_{t_p} -[P_{2p+2}, L]\big) \big|_{\ker (L-z)}
=\big( L_{t_p}+(L-z) P_{2p+2}\big) \big|_{\ker (L-z)}\notag\\
& = \bigg( a\bigg(\frac{ a_{t_p}}{a} -\frac{a_{t_p}^-}{a^-}
+2 (z-b^+) F_p^+
-2(z-b) F_p +G_{p+1}^+ -G_{p+1}^- \bigg) S^+\lb{3.2.22a} \\
& \;\;\; + \bigg( b_{t_p} +(z-b) \frac{a_{t_p}^-}{a^-}
+2(a^-)^2 F_p^- -2a^2 F_p^+  +(z-b) (G_{p+1}^- -G_{p+1})\bigg)\bigg)
\bigg|_{\ker(L-z)}.  \no
\end{align}
Hence one obtains
\begin{align}
& \frac{ a_{t_p}}{a} - \frac{ a_{t_p}^-}{a^-} = -2 (z-b^+) F_p^+
+2(z-b) F_p + G_{p+1}^- -G_{p+1}^+,
\lb{3.2.23time}\\
& b_{t_p} = -(z-b) \frac{a_{t_p}^-}{a^-} -2(a^-)^2 F_p^- +2a^2 F_p^+
-(z-b) \big(G_{p+1}^- -G_{p+1}\big). \lb{3.2.24time}
\end{align}
Further manipulations then yield,
\begin{align}
a_{t_p} &=-a\big(2(z-b^+) F_p^+
    +G_{p+1}^+ +G_{p+1}\big),  \lb{3.2.25time}  \\
b_{t_p} &=2 \big((z-b)^2 F_p + (z-b) G_{p+1} + a^2 F_p^+
- (a^-)^2 F_p^-\big).  \lb{3.2.26time}
\end{align}
Indeed, \eqref{3.2.25time} follows by adding $G_{p+1} -G_{p+1}$ to
\eqref{3.2.23time} (neglecting a  trivial summation constant), and an
insertion of \eqref{3.2.25time} into \eqref{3.2.24time} implies
\eqref{3.2.26time}. Varying $p\in\bbN_0$, equations \eqref{3.2.25time} and
\eqref{3.2.26time} provide an alternative description of the
time-dependent Toda hierarchy.

%%%%%%%%%%%%%%%%%%%%%%%%%%%%%%%%%%%%%%%%%%%
\begin{remark}\lb{r3.2.7}
{}{}From \eqref{1.2.4a}--\eqref{1.2.4c} and \eqref{1.2.11a}, \eqref{1.2.11b}
one concludes that the coefficient $a$ enters quadratically in
$F_p$ and $G_{p+1}$, and hence the Toda hierarchy \eqref{3.2.13}
$($respectively \eqref{3.2.29d}$)$ is invariant under the substitution
\begin{equation}
a \to a_\eps =\{\eps(n) a(n)\}_{n\in\bbZ},
\quad \eps (n) \in \{1, -1\}, \; n\in\bbZ. \lb{3.2.28}
\end{equation}
\end{remark} 
%%%%%%%%%%%%%%%%%%%%%%%%%%%%%%%%%%%%%%%%%%%

%%%%%%%%%%%%%%%%%%%%%%%%%%%%%%%%%%%%%
%%%%%%%%%%%%%%%%%%%%%%%%%%%%%%%%%%%%%
\section{Green's functions and high-energy expansions} \lb{s3}
%%%%%%%%%%%%%%%%%%%%%%%%%%%%%%%%%%%%%
%%%%%%%%%%%%%%%%%%%%%%%%%%%%%%%%%%%%%

In the following it will be necessary to associate certain operators to the difference expression $L$ in \eqref{1.2.2}. Throughout this section we assume that $a, b$ satisfy Hypothesis \ref{h2.1}.

We start with the Jacobi operator $H$ defined on $\ell^2(\bbZ)$ by
\begin{equation}
Hf = Lf, \quad f\in \dom(H)=\ell^2(\bbZ).  \lb{h9.1}
\end{equation}
In addition, we introduce the half-line Dirichlet operators $H^D_{+,n_0}$ on 
$\ell^2([n_0,\infty)\cap\bbZ)$ and $H^D_{-,n_0}$ on $\ell^2([n_0,-\infty)\cap\bbZ)$, 
$n_0\in\bbZ$, by 
\begin{align}
&\big(H^D_{+,n_0} u\big)(n) = \begin{cases} a(n_0)u(n_0+1)+b(n_0)u(n_0), & n=n_0, \\
(Lu)(n), & n\in [n_0+1,\infty)\cap\bbZ, \end{cases}  \no \\
&u \in  \dom(H^D_{+,n_0})=\ell^2([n_0,\infty)\cap\bbZ),  \lb{h9.2} \\
&\big(H^D_{-,n_0} v\big)(n) = \begin{cases} a(n_0)v(n_0+1)+b(n_0)v(n_0), & n=n_0, \\
(Lv)(n), & n\in (-\infty, n_0-1]\cap\bbZ, \end{cases}  \no \\
& v \in \dom(H^D_{-,n_0})=\ell^2((-\infty,n_0]\cap\bbZ).  \lb{h9.3}
\end{align}
(It is customary to join the Dirichlet-type boundary condition $u(n_0-1)=0$ in the case of 
$H^D_{+,n_0}$, and $v(n_0+1)=0$ in the case of $H^D_{-,n_0}$ in order to formally be able to avoid the case distinction in \eqref{h9.2} and \eqref{h9.3}.) The following elementary result about a spectral inclusion of $H$ and $H^D_{\pm,n_0}$ will be useful later.

%%%%%%%%%%%%%%%%%%%%%%%%%%%%%%%%%%%%%%%%%
\begin{theorem} \lb{ht9.1} 
Suppose $a,b$ satisfy Hypothesis \ref{h2.1}. Then the numerical range of $H$, and hence in particular, the spectrum of $H$ and $H^D_{\pm,n_0}$ is contained in the closed ball centered at the origin of radius $2\|a\|_\infty + \|b\|_\infty$, that is, 
\begin{equation}
\spec (H), \, \spec\big(H^D_{\pm,n_0}\big) \subseteq \ol{B(0;2\|a\|_\infty + \|b\|_\infty)}, \; 
n_0\in\bbZ.  \lb{h9.4}
\end{equation}
\end{theorem}
%%%%%%%%%%%%%%%%%%%%%%%%%%%%%%%%%%%%%%%%%
\begin{proof}
We denote by $W(T)$ the numerical range of a bounded linear operator 
$T\in\calB(\calH)$ in the complex, separable Hilbert space $\calH$,
\begin{equation}
W(T)=\{(g,Tg)_{\calH} \in\bbC \,|\, g\in\calH, \, \|g\|_{\calH} =1\}.   \lb{h9.5}
\end{equation}
It is well-known that $W(T)$ is convex and that its closure contains the spectrum of $T$, 
\begin{equation}
\spec(T) \subseteq \ol{W(T)}   \lb{h9.6}
\end{equation}
(see, e.g., \cite[Ch.\ 1]{GR97}). Elementary arguments then prove 
\begin{align}
& |(f,Hf)_{\ell^2(\bbZ)}| \leq (2\|a\|_\infty + \|b\|_\infty) \|f\|^2_{\ell^2(\bbZ)}, \quad 
f\in\ell^2(\bbZ),   \lb{h9.7}  \\
& \big|\big(w,H^D_{\pm,n_0}w\big)_{\ell^2([n_0,\pm\infty)\cap\bbZ)}\big|
 \leq (2\|a\|_\infty + \|b\|_\infty) 
\|w\|^2_{\ell^2([n_0,\pm\infty)\cap\bbZ)},  \lb{h9.8} \\  
& \hspace*{6.35cm} w\in\ell^2([n_0,\pm\infty)\cap\bbZ),   \no 
\end{align} 
and hence \eqref{h9.4}.
\end{proof}
%%%%%%%%%%%%%%%%%%%%%%%%%%%%%%%%%%%%%%%%%

Since $H$ is a bounded second-order difference operator in $\ell^2(\bbZ)$ with 
\begin{equation}
\| H \| \leq 2\|a\|_\infty + \|b\|_\infty,   \lb{h9.9}
\end{equation}
the resolvent $(H-zI)^{-1}$, $z\in\bbC\setminus\spec(H)$, of $H$ is thus a Carleman  integral operator (cf., e.g., \cite[Sect.\ 6.2]{We80}; the corresponding measures involved are of course discrete measures). In particular, introducing
\begin{align}
\psi_+(z,n)&=((H-zI)^{-1}\delta_0)(n), \quad n \geq 1, \; 
\quad z\in\bbC\setminus\spec(H),  \lb{h9.10} \\
\psi_-(z,n)&=((H-zI)^{-1}\delta_0)(n), \quad n \leq -1, \; 
\quad z\in\bbC\setminus\spec(H),  \lb{h9.11}
\end{align}
where
\begin{equation}
\delta_k (n)=\begin{cases} 1, & n=k, \\ 0, & n\in\bbZ\setminus\{k\}, 
\end{cases} \quad k\in\bbZ,  \lb{h9.12}
\end{equation}
and using the Jacobi equation $L\psi_\pm(z)=z\psi(z)$ to uniquely extend 
$\psi_\pm(z,n)$ to all $n\in [0,\mp\infty)\cap\bbZ$ (this is possible since by 
Hypothesis \ref{h2.1}, $a(n)\neq 0$ for all $n\in\bbZ$), one obtains,
\begin{equation}
L\psi_\pm(z)=z\psi_\pm(z), \quad 
\psi_\pm(z,\dott)\in\ell^2([n_0,\pm\infty)\cap\bbZ), \quad z\in\bbC\setminus\spec(H), 
 \; n_0\in\bbZ.  \lb{h2.5a}
\end{equation}
We note that the Weyl--Titchmarsh-type solutions $\psi_\pm(z,\dott)$ are unique up to normalization. Moreover, by the second inclusion in \eqref{h9.4}, 
\begin{equation}
\psi_\pm(z,\dott) \, \text{ is zero free for $|z|>(2\|a\|_\infty + \|b\|_\infty)$}  \lb{h9.13}
\end{equation}
since $\psi_\pm(z,n_0)=0$ for some $n_0\in\bbZ$ yields an eigenvalue of 
$H^D_{\pm,n_0\pm 1}$.

The Green's function $G(z,\dott,\dott)$ of $H$, that is, the integral kernel of the resolvent $(H-zI)^{-1}$ (with respect to discrete measures), is then given by
\begin{align}
\begin{split}
& G(z,m,n)=\frac{1}{W(\psi_-(z), \psi_+(z))}
\begin{cases}\psi_-(z,m) \psi_+(z,n), & m\le n, \\
 \psi_-(z,n) \psi_+(z,m), & m\geq n,  \end{cases}   \lb{h2.6a} \\
& \hspace*{5.15cm} z\in \bbC\setminus\spec(H), \; m, n \in\bbZ,  
\end{split}
\end{align}
where $W(f,g)(n)$ denotes the Wronskian
\begin{equation}
W(f,g)(n)= a(n)(f(n)g(n+1) - f(n+1)g(n)), \quad n\in\bbZ,      \lb{h2.7a}
\end{equation}
of complex-valued sequences 
$f=\{f(n)\}_{n\in\bbZ}$ and $g=\{g(n)\}_{n\in\bbZ} \subset \bbC$.

The corresponding Green's functions $G^D_{\pm,n_0}(z,\dott,\dott)$ of 
$H^D_{\pm,n_0}$, that is, the integral kernels of the resolvents $(H-zI)^{-1}$ (again with respect to discrete measures), are then given by
\begin{align}
& G^D_{+,n_0}(z,m,n)=\frac{-1}{a(n_0-1)\psi_+(z,n_0-1)}
\begin{cases}\phi_0(z,m,n_0-1) \psi_+(z,n), & m\le n, \\
 \phi_0(z,n,n_0-1) \psi_+(z,m), & m\geq n,  \end{cases}  \no \\
& \hspace*{6cm} z\in \bbC\setminus\spec(H^D_{+,n_0}), \; m, n \in\bbZ,   \lb{h9.14} \\
& G^D_{-,n_0}(z,m,n)=\frac{-1}{a(n_0+1)\psi_-(z,n_0+1)}
\begin{cases} \psi_-(z,m) \phi_0(z,n,n_0+1), & m\le n, \\
\psi_-(z,n)  \phi_0(z,m,n_0+1) , & m\geq n,  \end{cases}  \no  \\
& \hspace*{6cm} z\in \bbC\setminus\spec(H^D_{-,n_0}), \; m, n \in\bbZ,   \lb{h9.15}
\end{align}
where $\phi_0(z,\dott,n_0)$ is a solution of 
\begin{equation}
 L\psi =z\psi,  \lb{h2.2}
\end{equation}
satisfying the initial conditions
\begin{equation}
\phi_0(z,n_0,n_0)=0, \quad \phi_0(z,n_0+1,n_0)=1, \quad z\in\bbC.  \lb{h9.16}
\end{equation}

Next, assuming that $\psi$ satisfies \eqref{h2.2}, the function $\phi=\phi (z,n)$, introduced by
\begin{equation}
\phi=\frac{\psi^{+}}{\psi},  \lb{h2.4}
\end{equation}
satisfies the Riccati-type equation
\begin{equation}
a\phi +a^- (\phi^-)^{-1} +(b-z)=0. \lb{h2.5}
\end{equation}

Defining 
\begin{equation}
\phi_\pm(z,n)=\frac{\psi_\pm(z,n+1)}{\psi_\pm(z,n)}, \quad z\in\bbC\setminus 
\spec(H^D_{\pm,n\pm 1}), \; 
n\in\bbZ, \lb{h2.8a}
\end{equation}
and using the Green's function representations \eqref{h9.14}, \eqref{h9.15}, one computes
\begin{align}
\phi_+(z,n)&=-a(n)G^D_{+,n+1}(z,n+1,n+1)  \no  \\
&=-a(n)\big(\delta_{n+1},(H^D_{+,n+1}-zI)^{-1}) 
\delta_{n+1}\big)_{\ell^2([n+1,\infty)\cap\bbZ)},   \lb{h9.17} \\
& \hspace*{3.6cm}  z\in\bbC\setminus \spec(H^D_{+,n+1}), \; n\in\bbZ, \no
\end{align}
and
\begin{align} 
\phi_-(z,n)&=a(n)^{-1}\big(z-b(n)+a(n-1)^2 G^D_{-,n+1}(z,n-1,n-1)\big)  \no \\
&=a(n)^{-1}\Big(z-b(n)   \lb{h9.18}  \\
& \hspace*{1.65cm} +a(n-1)^2 \big(\delta_{n-1},(H^D_{+,n-1}-zI)^{-1}
\delta_{n-1}\big)_{\ell^2((-\infty,n-1]\cap\bbZ)}\Big),  \no \\
& \hspace*{5.95cm} z\in\bbC\setminus \spec(H^D_{-,n-1}), \; n\in\bbZ.  \no
\end{align}
We provide the following description of the asymptotic behavior of $\phi_\pm$ as $z\to\infty$.

%%%%%%%%%%%%%%%%%
\begin{lemma} \lb{l9.2}
Suppose $a,b$ satisfy Hypothesis \ref{h2.1}. Then $\phi_\pm$ has the following convergent expansion with respect to $1/z$ around $1/z=0$,
\begin{equation}
\phi_{\pm}(z) = 
\begin{cases}
a\sum_{j=1}^\infty \phi_{+,j}z^{-j},\\[1mm]
\frac1{a}\sum_{j=-1}^\infty \phi_{-,j}z^{-j},
\end{cases}\lb{h2.8}
\end{equation} 
where
\begin{align}
\phi_{+,1}&=1, \quad \phi_{+,2}=b^+,  \no \\ %\lb{h2.11}\\
\phi_{+,j+1}&= b^+ \phi_{+,j}+(a^+)^2\sum_{\ell=1}^{j-1} \phi^+_{+,j-\ell} \phi_{+,\ell}, 
\quad j\ge 2,\lb{h2.12}\\
\phi_{-,-1}&=1, \quad \phi_{-,0}=-b, \quad \phi_{-,1}=-(a^-)^2, \no \\ % \lb{h2.9} \\ 
\phi_{-,j+1}&= -b \phi^-_{-,j}-\sum_{\ell=0}^j \phi_{-,j-\ell} \phi^-_{-,\ell}, \quad j\ge 1.\lb{h2.10}
\end{align}
\end{lemma}
%%%%%%%%%%%%%%%%%
\begin{proof}
Since $H^D_{\pm,n\pm 1}$ are bounded operators, \eqref{h9.17} and \eqref{h9.18} prove the existence of an analytic expansion of $\phi_\pm(z)$ with respect to $1/z$ around $1/z=0$ for $|z|$ sufficiently large. Moreover, \eqref{h9.17} and 
\eqref{h9.18} yield the leading asymptotic behavior,
\begin{align}
\phi_+(z)\underset{z\to\infty}{=}&-a z^{-1}+ \Oh(z^{-2})\lb{h2.7}\\
\intertext{and}
\phi_-(z)\underset{z\to\infty}{=}&\frac{z}{a}+ \Oh(1).\lb{h2.6}
\end{align}
Thus, by making the ansatz \eqref{h2.8} for $\phi_\pm$ and inserting it into \eqref{h2.5} one finds \eqref{h2.12} and \eqref{h2.10}.
\end{proof}
%%%%%%%%%%%%%%%%%%%

For the record we note the following explicit expressions,
\begin{align}
\phi_{+,1}&=1, \no \\ 
\phi_{+,2}&=b^+, \no \\
 \phi_{+,3}&=(a^+)^2+(b^+)^2, \no \\
 \phi_{+,4}&=(a^+)^2(2b^+ +b^{++})+(b^+)^3,  \text{ etc.,}  \no \\ 
 \phi_{-,-1}&=1, \lb{h9.19} \\ 
 \phi_{-,0}&=-b, \no \\ 
 \phi_{-,1}&=-(a^-)^2,  \no \\
 \phi_{-,2}&=-(a^-)^2 b^-, \no \\ 
 \phi_{-,3}&=-(a^-)^2 ((a^{--})^2+(b^-)^2),  \text{ etc.}  \no 
\end{align}

Later on we will also need the convergent expansion of $\ln(\phi_+)$ with respect to 
$1/z$ for $1/|z|$ suficiently small and hence we note that
\begin{align}
\ln(\phi_+(z))& = \ln\bigg(a\sum_{j=1}^\infty \phi_{+,j}z^{-j}\bigg) \no \\
& = \ln\bigg(\frac{a}{z}\bigg)+\ln\bigg(1+\sum_{j=1}^\infty \phi_{+,j+1}z^{-j}\bigg)\no \\
& =  \ln\bigg(\frac{a}{z}\bigg)+ \sum_{j=1}^\infty \rho_{+,j} z^{-j}, 
\lb{h2.13}
\end{align}
where
\begin{equation}
\rho_{+,1}=\phi_{+,2}, \quad \rho_{+,j}=\phi_{+,j+1}-\sum_{\ell=1}^{j-1}\frac{\ell}{j}\, \phi_{+,j+1-\ell}\rho_{+,\ell},  \quad j\ge 2. \lb{h2.14}
\end{equation}
The first few explicitly read
\begin{align}
\rho_{+,1}&=b^{+},\no \\
\rho_{+,2}&= (a^{+})^2+\frac12 (b^{+})^2,  \lb{h9.20}  \\
\rho_{+,3}&= \frac13(b^{+})^3+ (a^{+})^2(b^{+}+ b^{++}), \text{ etc.}.  \no
\end{align}
Similarly, one finds for $1/|z|$ sufficiently small, 
\begin{align}
\ln(\phi_-(z))& = \ln\bigg(a^{-1}\sum_{j=-1}^\infty \phi_{-,j}z^{-j}\bigg) \no \\
& = \ln\bigg(\frac{z}{a}\bigg)+\ln\bigg(1+\sum_{j=1}^\infty \phi_{-,j-1}z^{-j}\bigg)\no \\
& =  \ln\bigg(\frac{z}{a}\bigg)+ \sum_{j=1}^\infty \rho_{-,j} z^{-j}, 
\lb{h2.13a}
\end{align}
where
\begin{equation}
\rho_{-,1}=\phi_{-,0}, \quad \rho_{-,j}=\phi_{-,j-1}-\sum_{\ell=1}^{j-1}\frac{\ell}{j}\, \phi_{-,j-1-\ell}\rho_{-,\ell}, 
\quad j\ge 2. \lb{h2.14a}
\end{equation}
The first few explicitly read
\begin{align}
\rho_{-,1}&=-b,\no \\
\rho_{-,2}&= - (a^{-})^2 - \frac12b^2,  \lb{h9.21}   \\
\rho_{-,3}&= -\frac13 b^3 - (a^{-})^2(b^- +b), \text{ etc.}  \no
\end{align}
Equations \eqref{h2.14} and \eqref{h2.14a} may be written as 
\begin{equation}
\rho_{\pm,1}=\phi_{\pm,1\pm1}, \quad \rho_{\pm,j}=\phi_{\pm,j\pm1}
-\sum_{\ell=1}^{j-1}\frac{\ell}{j}\, \phi_{\pm,j\pm1-\ell}\rho_{\pm,\ell}, 
\quad j\ge 2. \lb{h2.14aa}
\end{equation}

The next result shows that $\hat f_j$ and $\pm j \rho_{\pm,j}$ are equal up to terms that are total differences, that is, are of the form $(S^+ -I)d_{\pm,j}$ for some sequence 
$d_{\pm,j}$. The exact form of $d_{\pm,j}$ will not be needed later.

%%%%%%%%%%%%%%%%%%% 
%------------- lemma
\begin{lemma} \lb{lemmah2.4}
Suppose $a,b$ satisfy Hypothesis \ref{h2.1}. Then, 
\begin{equation}
\hat f_j=\pm j\rho_{\pm,j}+(S^+ - I) d_{\pm,j}, \quad j\in\bbN, \lb{h2.15}
\end{equation}
for some polynomials $d_{\pm,j}$, $j\in\bbN$, in $a$ and $b$ and certain shifts thereof.
\end{lemma}
%%%%%%%%%%%%%%%%%%%%
\begin{proof}
To shorten the notation we introduce the abbreviations 
\begin{equation}
g(z,n)=G(z,n,n), \quad h(z,n)=G(z,n,n+1), \quad (z,n)\in\bbC\times\bbZ.  \lb{h2.16}
\end{equation}
Thus, 
\begin{equation}
\phi_+=\frac{h}{g}.\lb{h2.17}
\end{equation}
To increase readability we suppress the display of variables in the subsequent computations. First, one observes that
\begin{align}
a g g^+ -h(ah-1)&=a\frac{\psi_+\psi_- \psi_+^+\psi_-^+}{W^2}
-\frac{\psi_+^+\psi_-}{W}\bigg(a\frac{\psi_+^+\psi_-}{W}-1 \bigg)\no \\
&= \frac{\psi_+^+\psi_-}{W^2}\big(a  \psi_+   \psi_-^+ - a \psi_+^+ \psi_ - +W \big) 
=0  \lb{h2.18}
\end{align}
and thus,
\begin{equation}
\phi_+^2= \frac{h^2}{g^2}= \frac{ah^2}{a g g^+}\, \frac{g^+}{g}
= \frac{ah}{ah -1}\,\frac{g^+}{g}.\lb{h2.19}
\end{equation}
This yields 
\begin{equation}
 \frac{d}{dz}\ln (\phi_+)= \frac12\frac{d}{dz}\ln\bigg( \frac{ah}{ah -1}\bigg)
 + \frac12\frac{d}{dz}\ln\bigg(\frac{g^+}{g}\bigg).\lb{h2.20}
\end{equation}
Next we study the first term on the right-hand side of \eqref{h2.20}, namely, 
\begin{equation}
\frac{d}{dz}\ln\bigg( \frac{ah}{ah -1}\bigg)
= \frac{h_z}{h}- \frac{ah_z}{ah-1}.     \lb{h2.21}
\end{equation}
We claim that
\begin{equation}
\frac{h_z}{h}- \frac{ah_z}{ah-1}
= (1-2ah)\bigg( \frac{g_z}{g}+\frac{g_z^+}{g^+} \bigg)+4ah_z.\lb{h2.22}
\end{equation}
To this end we first  take the logarithmic derivative of \eqref{h2.18}, that is, 
of $a g g^+ =h(ah-1)$, to obtain
\begin{equation}
\frac{g_z}{g}+\frac{g_z^+}{g^+}= \frac{h_z}{h}+  \frac{ah_z}{ah -1}.     \lb{h2.23}
\end{equation}
Thus,
\begin{align}
&\frac{h_z}{h}- \frac{ah_z}{ah-1}
-\bigg( (1-2ah)\bigg( \frac{g_z}{g}+\frac{g_z^+}{g^+} \bigg)+4ah_z\bigg) \no \\
&\quad = \frac{h_z}{h}- \frac{ah_z}{ah-1}
- (1-2ah)\bigg( \frac{h_z}{h}+  \frac{ah_z}{ah -1} \bigg)-4ah_z\no \\
&\quad=-\frac{2ah_z}{ah -1} \big(1+(ah -1)-ah\big)=0.    \lb{h2.24}
\end{align}
Combining \eqref{h2.20},  \eqref{h2.21},  and \eqref{h2.22} one finds
\begin{align}
\frac{d}{dz}\ln(\phi_+)&= \frac12\bigg((1-2ah)\bigg( \frac{g_z}{g}+\frac{g_z^+}{g^+}\bigg)+4ah_z\bigg)
+ \frac12\frac{d}{dz}\ln\bigg(\frac{g^+}{g}\bigg)  \no \\
&= \frac12(1-2ah)\bigg( \frac{g_z}{g}+\frac{g_z^+}{g^+}\bigg)
+2ah_z+ \frac12\frac{d}{dz}
\ln\bigg(\frac{g^+}{g}\bigg).     \lb{h2.25}
\end{align}
Considering the first term on the right-hand side of \eqref{h2.25} and subtracting $g$ 
then yields 
\begin{align}
&\frac12(1-2ah)\bigg( \frac{g_z}{g}+\frac{g_z^+}{g^+}\bigg)
+2ah_z-g  \no \\
&\quad = \frac12(1-2ah)\bigg(\frac{d}{dz}\ln(ah+a^-h^- -1)+
\frac{d}{dz}\ln(a^+h^+ +ah -1)\no \\
&\qquad-\frac{d}{dz}\ln\big((z-b^+)(z-b) \big)\bigg)
 +2ah_z-\frac{ah+a^-h^- -1}{z-b}\no \\
&\quad=\frac12(1-2ah)\bigg(\frac{ah_z+a^-h^-_z}{ah+a^-h^- -1}
+\frac{a^+h^+_z +ah_z}{a^+h^+ +ah -1}
 -\bigg(\frac{1}{z-b^+}+\frac{1}{z-b}\bigg)\bigg)\no \\
&\qquad +2ah_z-\frac{ah +a^-h^-}{z-b}+\frac{1}{z-b} \no \\
&\quad= \frac12\bigg(\frac{1}{z-b}-\frac{1}{z-b^+}\bigg)
+ \frac{ah}{z-b^+}-  \frac{a^-h^-}{z-b}\no \\
&\qquad+2ah_z+\frac12(1-2ah)\bigg(\frac{ah_z+a^-h^-_z}{ah+a^-h^- -1}
+\frac{a^+h^+_z +ah_z}{a^+h^+ +ah -1}\bigg),\lb{h2.26}
\end{align}
using 
\begin{equation}
(z-b)g-ah-a^-h^- +1=0.  \lb{h2.27}
\end{equation}
The latter follows from the definition \eqref{h2.16} of $g$ and $h$ (cf.\ \eqref{h2.6a}) and from  $L\psi_-=z\psi_-$. By purely algebraic manipulations one obtains that
\begin{align}
&2ah_z+\frac12(1-2ah)\bigg(\frac{ah_z+a^-h^-_z}{ah+a^-h^- -1}+
\frac{a^+h^+_z +ah_z}{a^+h^+ +ah -1}\bigg)\no \\
&\quad =\bigg(\frac{a^-h^- ah_z -a^-h^-_z ah}{ah +a^-h^- -1}
-\frac{ah a^+h_z^+ -ah_z a^+h^+}{a^+h^+ 
+ah -1} \bigg)\no \\
&\qquad +\frac12\bigg(\frac{a^+h^+_z- ah_z}{a^+h^+ +ah-1}
-\frac{ah_z-a^-h^-_z}{ah+a^-h^- -1}\bigg).\lb{h2.28}
\end{align}
Summarizing the computations thus far, one finds
\begin{align}
 \frac{d}{dz}\ln(\phi_+)&=g +\frac12\frac{d}{dz}\ln\bigg(\frac{g^+}{g}\bigg)
 +  \frac12\bigg(\frac{1}{z-b}-\frac{1}{z-b^+}\bigg)
+ \frac{ah}{z-b^+}-  \frac{a^-h^-}{z-b}\no \\
&\quad +\bigg(\frac{a^-h^- ah_z -a^-h^-_z ah}{ah +a^-h^- -1}
-\frac{ah a^+h_z^+ -ah_z a^+h^+}{a^+h^+ 
+ah -1} \bigg)\no \\
&\quad +\frac12\bigg(\frac{a^+h^+_z- ah_z}{a^+h^+ +ah-1}
-\frac{ah_z-a^-h^-_z}{ah+a^-h^- -1}\bigg)\no\\
&=g+(S^+ -  I)\Phi \lb{h2.29}
\end{align}
for some function $\Phi=\Phi(z,n)$.  Using  
\begin{equation}
g(z,n)=G(z,n,n) = -\sum_{j=0}^\infty \hat f_j(n) z^{-j-1}, \quad (z,n)\in\bbC\times\bbZ  
\lb{h2.30}
\end{equation}
for $1/|z|$ sufficiently small (see, e.g., \cite[Appendix C]{GH07}) and \eqref{h2.13}, one concludes
\begin{equation}
-z^{-1}-\sum_{j=1}^\infty j \rho_{+,j} z^{-j-1}=-\sum_{j=0}^\infty \hat f_j z^{-j-1}
+ (S^+ -I)\sum_{j=0}^\infty d_{+,j} z^{-j-1} \lb{h2.31}
\end{equation}
for some sequences $\{d_{+,j}(n)\}_{n\in\bbZ}$, $j\in\bbN_0$.

Noting that 
\begin{equation}
\phi_-=\frac{g^+}{h},\lb{h2.17a}
\end{equation}
one finds the analogous result regarding $\rho_{-,j}$ and $d_{-,j}$, $j\in\bbN$. 
\end{proof}
%-------------- end lemma

%%%%%%%%%%%%%%%%%%%%%%%%%%%%%%%%%%% 
\begin{remark}\lb{r3.2.7a}
Closely related to \eqref{h2.29} is the identity
\begin{equation}
\f{d}{dz} \ln\bigg(1-\f{1}{ah} \bigg)= -2 g +(S^+ - I)\bigg(-2\Phi+\f{d}{dz}\ln (g)\bigg),   
\lb{h2.17b}
\end{equation}
which follows from \eqref{h2.20} and \eqref{h2.29}, see also
equation $(3.19)$ in \cite{BG05}. Equation \eqref{h2.17b} is the analog of a well-known 
identity for the diagonal Green's function of Schr\"odinger operators, see, for instance, 
\cite[p.\ 369]{CL90}, \cite{GD75}, \cite[p.\ 99 and 122]{GH03}, and \cite{JM82}. 
\end{remark}
%%%%%%%%%%%%%%%%%%%%%%%%%%%%%%%%%%%% 

For later use we add the notion of degree of the quantities introduced.

%%%%%%%%%%%%%%%%%%%%%%%%%%%%%%%%%%% 
\begin{remark}\lb{r3.2.7aa}
Introduce
\begin{equation}
\deg (a^{(r)})=\deg (b^{(r)})=1, \quad r\in\bbZ. 
\end{equation}
This results in
\begin{equation}
\deg(\hat f_\ell)=\ell, \quad \deg(\hat g_\ell)=\ell+1, \quad \ell\in\bbN,
\end{equation}
using induction in the linear recursion relations 
\eqref{1.2.4a}--\eqref{1.2.4c}. Similarly, the recursion relations \eqref{h2.12}
and \eqref{h2.10} yield inductively that
\begin{equation}
\deg(\phi_{\pm,j\pm 1})=j, \quad j\in\bbN_0.
\end{equation}
\end{remark}
%%%%%%%%%%%%%%%%%%%%%%%%%%%%%%%%%%%% 

%%%%%%%%%%%%%%%%%%%%%%%%%%%%%%%%%%%%
%%%%%%%%%%%%%%%%%%%%%%%%%%%%%%%%%%%%
\section{Local conservation laws} \lb{s4}
%%%%%%%%%%%%%%%%%%%%%%%%%%%%%%%%%%%%
%%%%%%%%%%%%%%%%%%%%%%%%%%%%%%%%%%%%

Throughout this section (and with the only major exception of Theorem \ref{t9.6}) we make the following assumption:
%%%%%%%%%%%%%%%%%%%%%%%%%%%%%%%%%%%%% 
\begin{hypothesis} \lb{hh9.3}
Suppose that $a, b \colon \bbZ\times\bbR \to \bbC$ satisfy
\begin{align}
\begin{split}
& \sup_{(n,t_p)\in\bbN\times\bbR}\big(|a(n,t_p)|+|b(n,t_p)|\big) < \infty,  \lb{h9.22} \\
& a(n,\dott), \, b(n,\dott) \in C^1(\bbR), \; n\in\bbZ,  \quad 
 a(n,t_p)\neq 0 \, \text{ for all $(n,t_p)\in\bbZ\times\bbR$}.    
 \end{split}
\end{align}
\end{hypothesis}
%%%%%%%%%%%%%%%%%%%%%%%%%%%%%%%%%%%%% 

To simplify notations we denote the bounded difference operator defined on $\ell^2(\bbZ)$, generated by the finite difference expression $P_{2p+2}$ in 
\eqref{1.2.8}, by the same symbol $P_{2p+2}$.

We start with the following existence result.
%%%%%%%%%%%%%%%%%
\begin{theorem} \lb{t9.4} 
Let $(z,t_p)\in (\bbC\setminus\spec(H)) \times \bbR$ and suppose $a,b$ satisfy Hypothesis \ref{hh9.3} and $\TL_p(a,b)=0$ for some $p\in\bbN_0$.   
Then there exist Weyl--Titchmarsh-type solutions $\psi_\pm=\psi_\pm(z,n,t_p)$ such that 
\begin{equation}
\psi_\pm (z,\dott,t_p) \in\ell^2([n_0,\pm\infty)\cap\bbZ), \; n_0\in\bbZ,  
\quad \psi_\pm (z,n,\dott) \in C^1(\bbR),   \lb{h9.25} 
\end{equation}
and $\psi_\pm$ simultaneously satisfy the following two equations 
\begin{align}
& L(t_p)\psi_\pm(z,\dott,t_p)=z\psi_\pm(z,\dott,t_p),  \lb{h2.2A}  \\
& \psi_{\pm,t_p}(z,\dott,t_p)= P_{2p+2}(t_p)\psi_\pm(z,\dott,t_p), \lb{h2.1a}\\
& \quad =2a(t_p)F_p(z,\dott,t_p) \psi^{+}_{\pm}(z,\dott,t_p)+G_{\gg+1}(z,\dott,t_p) 
\psi_{\pm}(z,\dott,t_p).  \lb{h2.3}
\end{align}
Moreover, the Wronskian 
\begin{equation}
W(\psi_-(z,n,t_p),\psi_+(z,n,t_p)) \, 
\text{ is independent of $(n,t_p)\in\bbZ\times\bbR$.}   \lb{h9.26}
\end{equation}
\end{theorem}
%%%%%%%%%%%%%%%%%%
\begin{proof}
Applying $(H(t)-zI)^{-1}$ to $\delta_0$ (cf.\ \eqref{h9.10} and \eqref{h9.11}) yields the existence of Weyl--Titchmarsh-type solutions $\Psi_\pm$ of $L\psi=z\psi$ satisfying \eqref{h9.25}. Next, using the Lax commutator equation \eqref{3.2.11} one computes 
\begin{align}
\begin{split}
z\Psi_{\pm,t_p}&=(L \Psi_\pm)_{t_p}=L_{t_p}\Psi_\pm+L\Psi_{\pm,t_p}
=[P_{2p+2},L]\Psi_\pm + L \Psi_{\pm,t_p}  \\
&=zP_{2p+2}\Psi_\pm - L P_{2p+2}\Psi_\pm+L\Psi_{\pm,t_p}  \lb{h9.27}
\end{split}
\end{align}
and hence
\begin{equation}
(L-zI)(\Psi_{\pm,t_p} - P_{2p+2} \Psi_{\pm}) = 0.    \lb{h9.28}
\end{equation}
Thus, $\Psi_\pm$ satisfy
\begin{equation}
\Psi_{\pm,t_p} - P_{2p+2} \Psi_{\pm} = C_\pm \Psi_\pm + D_\pm \Psi_\mp.  \lb{h9.29}
\end{equation}
Introducing $\Psi_\pm = c_\pm \psi_\pm$, and choosing $c_\pm$ such that 
$c_{\pm,t_p} = C_\pm c_\pm$, one obtains
\begin{equation}
\psi_{\pm,t_p} - P_{2p+2} \psi_{\pm} = D_\pm \psi_\mp.  \lb{h9.30}
\end{equation}
Since $\psi_\pm \in\ell^2([n_0,\pm\infty)\cap\bbZ)$, $n_0\in\bbZ$, and 
$a, b$ satisfy Hypothesis \ref{hh9.3}, \eqref{1.2.13} shows that  
$P_{2p+2} \psi_\pm = (2a F_pS^+\psi_\pm + G_{p+1} \psi_\pm)
\in\ell^2([n_0,\pm\infty)\cap\bbZ)$. (Incidentally, this argument yields of course 
\eqref{h2.3}.) Moreover, since $\psi_\pm (z,n,t_p) 
= d_\pm (t_p) (H(t_p)-zI)^{-1} \delta_0)(n)$ for $n\in [\pm 1,\infty)\cap\bbZ$ and some 
$d_\pm \in C^1(\bbR)$, the calculation
\begin{equation}
\psi_{\pm,t_p}=d_{\pm,t_p} (H-zI)^{-1} \delta_0  
- d_\pm (H-zI)^{-1} H_{t_p} (H-zI)^{-1} \delta_0  
\lb{h9.31}
\end{equation}
also yields $\psi_{\pm,t_p}\in\ell^2([n_0,\pm\infty)\cap\bbZ)$. But then $D_\pm =0$ in \eqref{h9.30} since $\psi_\mp \notin \ell^2([n_0,\pm\infty)\cap\bbZ)$. This proves 
\eqref{h2.1a}. 

Since $\psi_\pm(z,\dott)$ satisfy $L\psi(z,\dott)=z\psi(z,\dott)$, the Wronskian 
\eqref{h9.26} is $n$-independent. To show also  its $t_p$-independence, a computation reveals 
\begin{equation}
\f{d}{d t_p} W(\psi_-,\psi_+)=\bigg(\f{a_{t_p}}{a} + 2 (z-b^+)F_p^+  + G_{p+1}^+ 
+ G_{p+1}\bigg) W(\psi_-,\psi_+) =0
\end{equation}  
using \eqref{h2.7a}, \eqref{h2.3}, and \eqref{3.2.25time}.
\end{proof}
%%%%%%%%%%%%%%%%%%

For the remainder of this section, $\psi_\pm$ will always refer to the Weyl--Titchmarsh solutions introduced in Theorem \ref{t9.4}.

\smallskip

The next result recalls the existence of a propagator $W_p$ associated with $P_{2p+2}$. 

%%%%%%%%%%%%%%%%%%
\begin{theorem} \lb{t9.5}
Assume Hypothesis \ref{hh9.3} and suppose $a,b$  satisfy $\TL_p(a,b)=0$ for some 
$p\in\bbN_0$. Then there is a propagator $W_p(s,t) \in \calB(\ell^2(\bbZ))$, $(s,t)\in\bbR^2$, satisfying
\begin{align}
& (i)\;\;\; W_p(t,t)=I, \quad t\in\bbR,  \lb{h9.32}  \\
& (ii)\;\; W_p(r,s)W_p(s,t)=W_p(r,t), \quad (r,s,t)\in\bbR^3,   \lb{h9.33}  \\
& (iii) \; W_p(s,t) \, \text{ is jointly strongly continuous in $(s,t)\in\bbR^2$,}  \lb{h9.34} 
\end{align}
such that for fixed $t_0\in\bbR$, $f_0\in\ell^2(\bbZ)$, 
\begin{equation}
f(t)=W_p(t,t_0)f_0, \quad  t\in\bbR,  \lb{h9.35}
\end{equation}
satisfies
\begin{equation}
\f{d}{dt} f(t) = P_{2p+2}(t) f(t), \quad f(t_0)=f_0.  \lb{h9.36}
\end{equation}
Moreover, $H(t)$ is similar to $H(s)$ for all $(s,t)\in\bbR^2$,
\begin{equation}
H(s) = W_p(s,t) H(t) W_p(s,t)^{-1}, \quad (s,t)\in\bbR^2.  \lb{h9.37}
\end{equation} 
This extends to appropriate functions of $H(t)$ and so, in particular, to its resolvent 
$(H(t)-zI)^{-1}$, $z\in \bbC\setminus \sigma(H(t))$, and hence also yields
\begin{equation}
\sigma (H(s)) = \sigma (H(t)), \quad (s,t)\in\bbR^2.  \lb{h9.38}
\end{equation}
\end{theorem}
%%%%%%%%%%%%%%%%%% 
\begin{proof}
\eqref{h9.32}--\eqref{h9.36} are standard results which follow, for instance, from Theorem X.69 of \cite{RS75} (under even weaker hypotheses on $a, b$). In particular, the propagator $W_p$ admits the norm convergent Dyson series
\begin{align}
& W_p(s,t)=I+\sum_{k\in\bbN} \int_s^t dt_1 \int_s^{t_1} dt_2 \cdots \int_s^{t_{k-1}} dt_k \, 
P_{2p+2}(t_1) P_{2p+2}(t_2) \cdots P_{2p+2}(t_k),  \no \\
& \hspace*{9.5cm} (s,t)\in\bbR^2.  \lb{h9.39}
\end{align}
Fixing $s\in\bbR$ and introducing the operator-valued function 
\begin{equation}
K(t)= W_p(s,t) H(t) W_p(s,t)^{-1}, \quad t\in\bbR, \lb{h9.40}
\end{equation}
one computes 
\begin{equation}
K' (t)f = W_p(s,t) (H'(t)-[P_{2p+2}(t),H(t)]) W_p(s,t)^{-1}f = 0, \quad t\in\bbR, \; 
f \in \ell^2(\bbZ),   \lb{h9.41}
\end{equation}
using the Lax commutator equation \eqref{3.2.11}. 
Thus, $K$ is independent of $t\in\bbR$ and $K(s)=H(s)$ then proves \eqref{h9.37}.
\end{proof}
%%%%%%%%%%%%%%%%%%

In the special case where $H(t)$, $t\in\bbR$, is self-adjoint, the operator $P_{2p+2}(t)$ is skew-adjoint, $P_{2p+2}(t)^*=-P_{2p+2}(t)$, $t\in\bbR$, and hence $W_p(s,t)$ is unitary 
for all $(s,t)\in\bbR^2$. 

Next we briefly recall the Toda initial value problem in a setting convenient for our purpose.

%%%%%%%%%%%%%%%%%%
\begin{theorem} \lb{t9.6} 
Let $t_{0,p}\in\bbR$ and suppose $a^{(0)}, b^{(0)} \in \ell^\infty(\bbZ)$. Then the $p$th Toda lattice initial value problem
\begin{equation}
\TL_p(a,b)=0, \quad (a,b)\big|_{t_p=t_{0,p}} = \big(a^{(0)},b^{(0)}\big)  \lb{h9.42}
\end{equation}
for some $p\in\bbN_0$ has a unique, local, and  smooth solution in time, that is, there exists a $T_0>0$ such that
\begin{equation}
a(\dott), \, b(\dott) \in C^\infty((t_{0,p}-T_0,t_{0,p}+T_0),\ell^\infty(\bbZ)).  \lb{h9.43}
\end{equation}
\end{theorem}
%%%%%%%%%%%%%%%%%%
\begin{proof}
This follows from standard results in \cite[Sect.\ 4.1]{AMR88} and has been exploited in the self-adjoint case in \cite[Sect.\ 12.2]{Te00}. More precisely, local existence and uniqueness as well as smoothness of the solution of the initial value problem 
\eqref{h9.42} (cf.\ also \eqref{3.2.13}) follows from \cite[Theorem\ 4.1.5]{AMR88} since $f_{p+1}$ and $g_{p+1}$ depend only polynomially on $a, b$ and certain of their shifts,  and the fact that the Toda flows are autonomous. 
\end{proof}
%%%%%%%%%%%%%%%%%%

%%%%%%%%%%%%%%%%%%
\begin{remark} \lb{r9.7} 
$(i)$ Theorem\ 4.1.5 in \cite{AMR88} also shows that Theorem \ref{t9.6} extends to the case where $a^{(0)}, b^{(0)} \in \ell^\infty(\bbZ)$ is replaced, for instance, by 
\begin{equation}
\big\{a^{(0)}(n)^2-\tfrac{1}{4}\big\}_{n\in\bbZ}, 
\big\{b^{(0)}(n)\big\}_{n\in\bbZ} \in\ell^1(\bbZ).  \lb{h9.43a}
\end{equation}
To apply the Banach space setting of \cite[Theorem\ 4.1.5]{AMR88} in this situation, one introduces a new sequence 
\begin{equation}
\big\{c^{(0)}(n)\big\}_{n\in\bbZ}=\big\{a^{(0)}(n)^2-\tfrac{1}{4}\big\}_{n\in\bbZ},   
\lb{h9.43b}
\end{equation}
and then demands
\begin{equation}
\big\{b^{(0)}(n)\big\}_{n\in\bbZ}, 
\big\{c^{(0)}(n)\big\}_{n\in\bbZ}  \in\ell^1(\bbZ),  \lb{h9.43c}
\end{equation}
upon substituting $\big\{a^{(0)}(n)^2\big\}_{n\in\bbZ}$ by 
$\big\{c^{(0)}(n)+\tfrac{1}{4}\big\}_{n\in\bbZ}$ in the equations of the Toda hierarchy. The latter is possible by Remark \ref{r3.2.7} since only $a^2$ but not $a$ itself enters the Toda hierarchy of evolution equations. This observation will be used in the proof of Theorem \ref{ht5.6}. \\ 
$(ii)$ In the special case where $H(t)$, $t\in\bbR$, is self-adjoint, one obtains 
\begin{equation}
\sup_{(n,t_p)\in\bbN\times(t_{0,p}-T_0,t_{0,p}+T_0)}\big(|a(n,t_p)|+|b(n,t_p)|\big) 
\leq 2 \|H(t_p)\| = 2 \|H(t_{0,p})\|,   \lb{h9.44}
\end{equation}
using $($the local version of\,$)$ the isospectral property \eqref{h9.38} of Toda flows. A further application of \cite[Proposition 4.1.22]{AMR88} then yields a unique, global, and smooth solution of  the $p$th Toda lattice initial value problem \eqref{h9.42}.  Moreover, the same argument shows that if  $a,b$ satisfy Hypothesis \ref{hh9.3} and  the $p$th Toda equation $\TL_p(a,b)=0$, then $a, b$ are actually smooth with respect to 
$t_p\in\bbR$, that is, 
\begin{equation}
a(n,\dott), \, b(n,\dott) \in C^\infty(\bbR),  \quad  n\in\bbZ.  \lb{h9.45}
\end{equation} 
\end{remark}
%%%%%%%%%%%%%%%%%%

%%%%%%%%%%%%%%%%%%
%-------------------- theorem
\begin{theorem} \lb{thm3.1}
Assume Hypothesis \ref{hh9.3} and suppose $a,b$ satisfy    
$\TL_p(a,b)=0$ for some $p\in\bbN_0$. Then the following infinite sequence of local conservation laws holds,
\begin{align}
\partial_{t_p} \ln (a) - (S^+ -I) f_{p+1} &=0, \lb{h3.1}\\
\partial_{t_p} \rho_{\pm,j} + (S^+ - I)J_{\pm,p, j} &=0, 
\quad j\in \bbN,  \lb{h3.2}
\end{align}
where 
\begin{equation}
J_{\pm,p,j} = -2\bigg(a^2\sum_{\ell=0}^p f_\ell \phi_{\pm,p+j-\ell} \bigg),   \lb{h3.3}
\end{equation}
and $\phi_{\pm,j}$, $j\in\bbN$, are given by \eqref{h2.12} and \eqref{h2.10}, and 
$\rho_{\pm,j}$, $j\in\bbN$,  are given by \eqref{h2.14} and \eqref{h2.14a}.
\end{theorem}
%%%%%%%%%%%%%%%%%
\begin{proof}
Using \eqref{h2.3} and \eqref{h2.4} one computes
\begin{align}
\partial_{t_p} \ln\bigg(\frac{\psi^+_+}{\psi_+}\bigg)&= \frac{\psi^+_{+,t_p}}{\psi^+_+}
-\frac{\psi_{+,t_p}}{\psi_+}\no \\
% &= 2a^+ F_p^+\frac{\psi^{++}_+}{\psi^+_+}+ G_{p+1+^+
 %    -2a  F_p \frac{\psi^{+}_+}{\psi_+} - G_{p+1} \no\\
&= 2a^+ F_p^+\phi^+_+ -2a  F_p\phi_+ + G_{p+1}^+- G_{p+1}\no \\
     &= (S^+ -I)\big(2a  F_p\phi_+ +  G_{p+1}\big).\lb{h3.4}
\end{align}
Utilizing the expansion \eqref{h2.8} of  $\phi_+$ as $z\to\infty$ one finds
\begin{align}
2a  F_p\phi_+ &+  G_{p+1} = 2a^2 \bigg(\sum_{\ell=0}^p f_{p-\ell}z^\ell\bigg)
\bigg( \sum_{j=1}^\infty \phi_{+,j}z^{-j}\bigg)
-z^{p+1}+\sum_{\ell=0}^p g_{p-\ell} z^\ell +f_{p+1} \no\\
&=\bigg(2a^2 \sum_{\ell=1}^p f_{p-\ell}\phi_{+,\ell}+ g_p+f_{p+1}\bigg)
+\sum_{k=1}^p\bigg(2a^2 \sum_{\ell=1}^p  f_{p-\ell}\phi_{+,\ell}+g_{p-k}\bigg)z^k\no \\
&\quad -z^{p+1}+2a^2 \sum_{j=1}^\infty\bigg(\sum_{\ell=0}^p f_{\ell}\phi_{+,p+j-\ell}\bigg)z^{-j}.\lb{h3.5}
\end{align}
On the other hand, using \eqref{h2.4} and the expansions \eqref{h2.8} and 
\eqref{h2.13}, one concludes
\begin{align}
\partial_{t_p} \ln\bigg(\frac{\psi^+_+}{\psi_+}\bigg)=\partial_{t_p} \ln(\phi_+)
= \frac{a_{t_p}}{a}+ \sum_{j=1}^{\infty} (\partial_{t_p} \rho_{+,j}) z^{-j}. \lb{h3.6}
\end{align}
Combining these equations one infers 
\begin{align}
\frac{a_{t_p}}{a}+ \sum_{j=1}^{\infty} (\partial_{t_p} \rho_{+,j}) z^{-j}
&=(S^+ -I)\bigg(2a^2 \sum_{\ell=1}^p  f_{p-\ell}\phi_{+,\ell}+ g_p+f_{p+1}\no \\
&\qquad\qquad\qquad+\sum_{k=1}^p\bigg(2a^2 \sum_{\ell=k+1}^p  f_{p-\ell}\phi_{+,\ell-k} +g_{p-k}\bigg)z^k   \no \\
&\qquad\qquad\qquad  -z^{p+1}
+2a^2 \sum_{j=1}^\infty\sum_{\ell=0}^p  f_{\ell}\phi_{+,p+j-\ell}z^{-j} \bigg)\no\\
&= (S^+ -I)\bigg(2a^2 \sum_{\ell=1}^p  f_{p-\ell}\phi_{+,\ell}+ g_p+f_{p+1}\bigg)\no \\
&\quad +(S^+ -I) \bigg(2a^2 \sum_{j=1}^\infty\sum_{\ell=0}^p  f_{\ell}\phi_{+,p+j-\ell}z^{-j}\bigg)    \lb{h3.7} \\
&= (S^+ -I)f_{p+1}
+(S^+ -I) \bigg(2a^2 \sum_{j=1}^\infty\sum_{\ell=0}^p  f_{\ell}\phi_{+,p+j-\ell}z^{-j}\bigg) 
\no
\end{align}
and hence \eqref{h3.2}, \eqref{h3.3} in the case of $\rho_{+,j}$. Here we used the first of the equations in $\TL_p(a,b)=0$ as well as the fact that the left-hand side of \eqref{h3.7} contains no positive powers of $z$. 

Similarly, one can start with $\phi_-$ and finds
\begin{align}
\partial_{t_p} \ln\bigg(\frac{\psi^+_-}{\psi_-}\bigg)&=  (S^+ -I)\big(2a  F_p\phi_- 
+  G_{p+1}\big)\no\\
&=(S^+ -I) \bigg( 2\phi_{-,0}f_p+2\sum_{\ell=0}^{p-1} f_\ell\phi_{-,p-\ell}+g_p+f_{p+1} \bigg)\no \\
&\quad +(S^+ -I)\sum_{j=1}^{p-1}\bigg(2\sum_{\ell=0}^{p-1-j}f_\ell \phi_{-,p-j-\ell}
+2\phi_{-,-1}f_{p-j+1}\no \\
&\hspace*{5.4cm} +2\phi_{-,0}f_{p-j}+g_{p-j} \bigg)z^j  \no \\
&\quad +(S^+ -I)\big((2\phi_{-,-1}f_{p-1}+2\phi_{-,0}f_{0}+g_0)z^p 
 +(2\phi_{-,-1}f_{0}-1)z^{p+1}\big) \no\\
&\quad+(S^+ -I)\sum_{j=1}^{\infty}\bigg(2\sum_{\ell=0}^{p}f_\ell 
\phi_{-,p+j-\ell}\bigg) z^{-j}.\lb{h3.5a}
\end{align}
Thus,
\begin{align}
- \frac{a_{t_p}}{a}+ \sum_{j=1}^{\infty}(\partial_{t_p} \rho_{-,j}) z^{-j}&= 
\partial_{t_p} \ln(\phi_+)
=  \partial_{t_p} \ln\bigg(\frac{\psi^+_+}{\psi_+}\bigg)\no \\
& =(S^+ -I)\bigg( 2\phi_{-,0}f_p+2\sum_{\ell=0}^{p-1} f_\ell\phi_{-,p-\ell}+g_p+f_{p+1} \bigg)\no \\
&\quad+(S^+ -I)\sum_{j=1}^{\infty}\bigg(2\sum_{\ell=0}^{p}f_\ell \phi_{-,p+j-\ell}\bigg) 
z^{-j},      \lb{h3.5aa}
\end{align}
implying 
\begin{equation}
\partial_{t_p} \rho_{-,j}=(S^+ -I)\bigg(2\sum_{\ell=0}^{p}f_\ell \phi_{-,p+j-\ell}\bigg), 
\quad j\in\bbN  \lb{h3.5ab}
\end{equation}
and hence \eqref{h3.2}, \eqref{h3.3} in the case of $\rho_{-,j}$.
\end{proof}
%------------- end theorem
%%%%%%%%%%%%%%%%%

%%%%%%%%%%%%%%%%%
%------- remark
\begin{remark}  \lb{rem4.2}
We emphasize that the sequence \eqref{h3.5ab} yields no new conservation laws, and so the latter is equivalent to that in \eqref{h3.7}. \\
The first local conservation law, \eqref{h3.1}, is of course nothing but the first equation in $\TL_p(a,b)=0$, namely $a_{t_p}=a(f_{p+1}^+ - f_{p+1})$. The second equation in  $\TL_p(a,b)=0$ is \eqref{h3.2} for $j=1$, 
namely\footnote{Using the second equation in  $\TL_p(a,b)=0$ one infers that $2a^2\sum_{\ell=0}^p  \hat f_{\ell}\phi_{+,p+1-\ell}= -\hat g_{p+1}$ up to a constant, that is, an element in the kernel of $S^+-I$. Using the notion of a degree (cf.\ Remark \ref{r3.2.7aa}), one concludes that the constant equals zero.}  $b^+_{t_p}= -(S^+-I)g_{p+1}$. 
The first few local conservation laws explicitly read as follows: \\
$(i)$ $p=0$:
\begin{equation}
\partial_{t_0} \rho_{+,j}= 2(S^+ -I)\phi_{+,j},
\end{equation}
in particular, 
\begin{align}
j&=1: \qquad  \partial_{t_0} b^+ = 2(S^+ -I)a^2 , \no \\
j&=2:  \qquad  \partial_{t_0}\bigg((a^{+})^2+\frac12(b^{+})^2\bigg)= 2(S^+ -I)a^2b^{+}. \no
\end{align}
$(ii)$ $p=1$: 
\begin{equation}
\partial_{t_1} \rho_{+,j}= 2(S^+ -I)(\phi_{+,j+1}+(b+c_1)\phi_{+,j}),
\end{equation}
in particular, 
\begin{align}
j&=1: \qquad  \partial_{t_1} b^+ =2(S^+ -I)\big(a^2\big(b^{+}+b\big)+c_1\big), \no \\
j&=2:  \qquad  \partial_{t_1}\bigg((a^{+})^2+\frac12(b^{+})^2\bigg)
= 2(S^+ -I)\big( a^2\big((a^{+})^2+b^+(b+b^+)\big)+c_1b^+\big).\no
\end{align}
%\begin{align}
%-\partial_{t_p}b^+&=2(S^+-I)  \big(a^2 \sum_{\ell=0}^p  f_{\ell}\phi_{-,p+1-\ell}\big)\\
%& \dots
%     \lb{h3.8}
%\end{align}
Using Lemma \ref{lemmah2.4}, one observes that one can replace $\rho_{\pm,j}$ in \eqref{h3.2} by $\hat f_j$ by suitably adjusting the right-hand side. 

An obvious consequence of the local conservation laws is that, assuming sufficient decay of the sequences $a^2-\frac14$ and $b$, one obtains 
\begin{equation}
\frac{d}{dt_p}\sum_{n\in\bbZ} \rho_{\pm,j}(n,t_p)=\frac{d}{dt_p}\sum_{n\in\bbZ} \hat f_j(n,t_p)=0. 
\end{equation}
\end{remark}
%%%%%%%%%%%%%%%%%%

%%%%%%%%%%%%%%%%%%%%%%%%%%%%%%%%%%%%%
%%%%%%%%%%%%%%%%%%%%%%%%%%%%%%%%%%%%%
\section{Hamiltonian formalism, variational derivatives} \lb{s5}
%%%%%%%%%%%%%%%%%%%%%%%%%%%%%%%%%%%%%
%%%%%%%%%%%%%%%%%%%%%%%%%%%%%%%%%%%%%

We start this section by a short review of variational derivatives for discrete systems. 
Consider the functional
\begin{align}
\begin{split}
& \calF\colon \funksj^\kappa\to \bbC,   \lb{h5.1A} \\
& \calF(u)=\sum_{n\in\bbZ} 
F\big(n,u(n), u^{(+1)}(n), u^{(-1)}(n), \dots, u^{(k)}(n), u^{(-k)}(n)\big)
\end{split}
\end{align}
for some $\kappa\in\bbN$ and $k\in\bbN_0$, where $F\colon \bbZ\times \bbC^{2r\kappa} \to \bbC$ is $C^1$ with respect to the $2r\kappa$ complex-valued entries and where
\begin{equation}
u^{(s)}= S^{(s)} u, \quad
 S^{(s)}= \begin{cases} (S^+)^s u & \text{if $s\ge0$}, \\
(S^{-})^{-s} u & \text{if $s<0$}, \end{cases} \quad u \in \ell^\infty(\bbZ)^\kappa.  
\lb{h5.2A}
\end{equation}
For brevity we  write 
\begin{equation}
F(n,u)=F\big(n,u(n), u^{(+1)}(n), u^{(-1)}(n), \dots, u^{(k)}(n), u^{(-k)}(n)\big),  
\lb{h5.3A}
\end{equation}
and it is assumed that  $\{F(n,u)\}_{n\in\bbZ}\in\ell^1(\bbZ)$.

For any $v\in \ell^1(\bbZ)^\kappa$ one computes for the differential $d\calF$
\begin{align}
(d\calF)_u (v)&=\frac{d}{d\epsilon}\calF(u+\epsilon v)\big|_{\epsilon=0} \no \\
&=\sum_{n\in\bbZ}\bigg(\frac{\partial F(n,u)}{\partial u} v(n)+ \frac{\partial F(n,u)}{\partial u^{(+1)}} v^{(+1)}(n)
+\frac{\partial F(n,u)}{\partial u^{(-1)}} v^{(-1)}(n) \no \\
&\hspace*{1.3cm}  +\cdots+ \frac{\partial F(n,u)}{\partial u^{(k)}} v^{(k)}(n)
+\frac{\partial F(n,u)}{\partial u^{(-k)}} v^{(-k)}(n) \bigg)\no \\
&= \sum_{n\in\bbZ}\bigg(\frac{\partial F(n,u)}{\partial u}+ S^{(-1)}\frac{\partial F(n,u)}{\partial u^{(+1)}} + 
S^{(+1)}\frac{\partial F(n,u)}{\partial u^{(-1)}}\no \\
&\hspace*{1.3cm}  +\cdots+S^{(-k)}\frac{\partial F(n,u)}{\partial u^{(k)}} 
+S^{(k)}\frac{\partial F(n,u)}{\partial u^{(-k)}} \bigg)v(n),   \lb{h5.4a}
\end{align}
assuming
\begin{align}
\begin{split}
& \{F(n,u)\}_{n\in\bbZ}, \, 
\bigg\{\frac{\partial F(n,u)}{\partial u^{(\pm j)}}\bigg\}_{n\in\bbZ} \in \ell^1(\bbZ),  
\quad j=1,\dots,k.   \lb{h5.5a}
\end{split}
\end{align}

Because of the result \eqref{h5.4a}, we thus introduce the gradient and the variational derivative of $\calF$ by
\begin{align}
(\nabla\calF)_u&=\frac{\delta F}{\delta u}   \lb{hA.5}  \\
&= \frac{\partial F}{\partial u}+ S^{(-1)}\frac{\partial F}{\partial u^{(+1)}} + 
S^{(+1)}\frac{\partial F}{\partial u^{(-1)}}+\cdots
+S^{(-k)}\frac{\partial F}{\partial u^{(k)}} 
+S^{(k)}\frac{\partial F}{\partial u^{(-k)}}, \no
\end{align} 
assuming \eqref{h5.5a}.

To establish the connection with the Toda hierarchy we make the following assumption for the remainder of this section. 
%---- hypothesis
\begin{hypothesis} \lb{hyp5.1}
Suppose
\begin{equation}
a, \,  a^{-1},\, b \in \ell^\infty (\bbZ).   \lb{h5.2.1}
\end{equation}
\end{hypothesis}
%----- end hypothesis

We introduce the difference expressions
\begin{align}
\begin{split}
D&=\begin{pmatrix} 0& D_1\\D_2 & 0\end{pmatrix}, \quad  D_1=aS^+ -a, \; D_2=a- a^- S^-,   \lb{h5.1} \\
D^{-1} &=\begin{pmatrix} 0& D_2^{-1}\\ D_1^{-1} & 0\end{pmatrix}, \quad 
(D^{-1})^\dagger=\begin{pmatrix} 0& (D_1^{-1})^\dagger\\ (D_2^{-1})^\dagger & 0\end{pmatrix}, 
\end{split}
\end{align}
where
\begin{align}
(D_1^{-1}u)(n)&= \sum_{m=-\infty}^{n-1}\frac{u(m)}{a(m)}, \quad
(D_2^{-1}u)(n)= \frac{1}{a(n)}\sum_{m=-\infty}^{n}u(m),  \lb{h5.2}  \\
((D_1^{-1})^\dagger u)(n)&=\frac{1}{a(n)}\sum_{m=n+1}^{\infty}u(m) , 
\quad ((D_2^{-1})^\dagger u)(n)=   \sum_{m=n}^{\infty}\frac{u(m)}{a(m)},  
\quad u \in \ell^1(\bbZ).  \no 
\end{align}
Viewing $D, D^{-1}$ and $D_1, D_1^{-1}, D_2, D_2^{-1}$ as operators on $\ell^1(\bbZ)^2$  and $\ell^1(\bbZ)$, respectively, one concludes that 
\begin{align}
D D^{-1}=I_{\ell^1(\bbZ)^2}, \quad D_1D_1^{-1}=I_{\ell^1(\bbZ)}, \quad 
D_2D_2^{-1}=I_{\ell^1(\bbZ)}. \lb{h5.3}
\end{align}

Next, let $\calF$ be a functional of the type 
\begin{align}
&\calF\colon \ell^\infty(\bbZ)^2 \to\bbC,   \no \\
& \calF(a,b) = \sum_{n\in\bbZ} F(n,a,b, a^{(+1)}, b^{(+1)},a^{(-1)}, b^{(-1)},\dots, a^{(+k)}, b^{(+k)},a^{(-k)}, b^{(-k)})  \no \\
& \qquad\quad  = \sum_{n\in\bbZ} F(n,a,b),   \lb{h5.3a} 
\end{align}
assuming
\begin{align}
& \{F(n,a,b)\}_{n\in\bbZ}, \, 
\bigg\{\frac{\partial F(n,a,b)}{\partial a^{(\pm j)}}\bigg\}_{n\in\bbZ}, \, 
\bigg\{\frac{\partial F(n,a,b)}{\partial b^{(\pm j)}}\bigg\}_{n\in\bbZ} \in \ell^1(\bbZ),   
\quad  j=1,\dots,k.   \lb{h5.5aa}
\end{align}
The gradient $\nabla \calF$ and symplectic gradient $\nabla_s \calF$ of $\calF$ are then defined by
\begin{equation}
(\nabla \calF)_{a,b}=\begin{pmatrix} (\nabla \calF)_a \\ (\nabla \calF)_b \end{pmatrix} 
= \begin{pmatrix}  \f{\delta \calF}{\delta a} \\[1.5mm]  \f{\delta \calF}{\delta b} 
\end{pmatrix}
\end{equation}
and
\begin{equation}
(\nabla_s \calF)_{a,b}= D(\nabla\calF)_{a,b}
= D\begin{pmatrix} 
(\nabla\calF)_{a}\\ (\nabla\calF)_{b}\end{pmatrix},   \lb{h5.4}
\end{equation}
respectively. In addition, we introduce the bilinear form 
\begin{align}
&\Omega\colon \ell^1(\bbZ)^2\times \ell^1(\bbZ)^2\to \bbC, \no \\
&\Omega(u,v)=\frac12 \sum_{n\in\bbZ}\Big( (D^{-1} u)(n) \cdot v(n)+ u(n)\cdot ((D^{-1})^\dagger v)(n)\Big). \lb{h5.5}
\end{align}
One then concludes that
\begin{align}
\begin{split}
\Omega(Du,v)&=\sum_{n\in\bbZ} u(n) \cdot v(n)
=\sum_{n\in\bbZ} \big(u_1(n)v_1(n)+u_2(n)v_2(n)\big) \\
&=\langle u, v\rangle_{\ell^2(\bbZ)^2}, \quad u, v \in \ell^1(\bbZ)^2,  \lb{h5.6}
\end{split}
\end{align}
where $\langle \dott,\dott \rangle_{\ell^2(\bbZ)^2}$ denotes the ``real'' inner product in $ \ell^2(\bbZ)^2$, that is,
\begin{align}
\begin{split}
& \langle \dott,\dott \rangle_{\ell^2(\bbZ)^2} \colon \ell^2(\bbZ)^2\times\ell^2(\bbZ)^2 \to \bbC, \\
&\langle u, v \rangle_{\ell^2(\bbZ)^2} =\sum_{n\in\bbZ} u(n)\cdot v(n) 
=\sum_{n\in\bbZ} \big(u_1(n)v_1(n)+u_2(n)v_2(n)\big).
\end{split}
\end{align}
In addition, one obtains 
\begin{equation}
(d\calF)_{a,b}(v)=\langle (\nabla\calF)_{a,b},v\rangle_{\ell^2(\bbZ)^2}
=\Omega(D(\nabla \calF)_{a,b},v) =\Omega((\nabla_s \calF)_{a,b},v). \lb{h5.7}
\end{equation}
Given two functionals $\calF_1,\calF_2$ we define their Poisson bracket by
\begin{align}
\{\calF_1,\calF_2\}&=d\calF_1(\nabla_s\calF_2)= \Omega(\nabla_s\calF_1,\nabla_s\calF_2)\no \\
&= \Omega(D\nabla\calF_1,D\nabla\calF_2) =\langle \nabla\calF_1,D\nabla\calF_2\rangle_{\ell^2(\bbZ)^2} \no \\
&=\sum_{n\in\bbZ}
\begin{pmatrix}\frac{\delta F_1}{\delta a}(n)\\[1.5mm] \frac{\delta F_1}{\delta b}(n)\end{pmatrix} \cdot 
D\begin{pmatrix}\frac{\delta F_2}{\delta a}(n)\\[1.5mm] \frac{\delta F_2}{\delta b}(n)\end{pmatrix}.  \lb{h5.8}
\end{align}
One then verifies that both the Jacobi identity 
\begin{equation}
\{\{\calF_1,\calF_2\},\calF_3\}+ \{\{\calF_2,\calF_3\},\calF_1\}+\{\{\calF_3,\calF_1\},\calF_2\}=0, \lb{h5.9}
\end{equation}
as well as the Leibniz rule
\begin{equation}
\{\calF_1,\calF_2\calF_3\}= \{\calF_1,\calF_2\}\calF_3+ \calF_2\{\calF_1,\calF_3\},  
\lb{h5.10}
\end{equation}
hold. 

If $\calF$ is a smooth functional and $(a,b)$ develops according to a Hamiltonian flow  with Hamiltonian $\calH$, that is,
\begin{equation}
\begin{pmatrix} a\\ b\end{pmatrix}_t=(\nabla_s\calH)_{a,b}=D(\nabla\calH)_{a,b}
%=D \frac{\delta H}{\delta u}(n) 
=D\begin{pmatrix} \frac{\delta H}{\delta a}\\[1mm] \frac{\delta H}{\delta b}\end{pmatrix}, \lb{h5.11}
\end{equation}
then
\begin{align}
\frac{d\calF}{dt}&=\frac{d}{dt}\sum_{n\in\bbZ} F(n,a,b) \no \\
%F\big(n,u(n), u^{(+1)}(n), u^{(-1)}(n), \dots, u^{(r)}(n), u^{(-r)}(n)\big) \no \\
&= \sum_{n\in\bbZ}  \begin{pmatrix} \frac{\delta F}{\delta a}(n)\\[1mm] \frac{\delta F}{\delta b}(n)\end{pmatrix}\cdot \begin{pmatrix} a(n)\\ b(n)\end{pmatrix}_t 
=  \sum_{n\in\bbZ}  \begin{pmatrix} \frac{\delta F}{\delta a}(n)\\[1mm] \frac{\delta F}{\delta b}(n)\end{pmatrix} \cdot 
  D \begin{pmatrix} \frac{\delta H}{\delta a}(n)\\[1mm] \frac{\delta H}{\delta b}(n)\end{pmatrix} \no \\
&= \{\calF,\calH\}. \lb{h5.12}
\end{align}
Here, and in the remainder of this section and the next, time-dependent equations  such as \eqref{h5.12} are viewed locally in time, that is, assumed to hold on some open 
$t$-interval $\bbI\subseteq\bbR$.
 
If a functional $\calG$ is in involution with the Hamiltonian $\calH$, that is,
\begin{equation}
\{\calG,\calH\}=0, \lb{h5.13}
\end{equation}
then it is conserved, that is,
\begin{equation}
\frac{d\calG}{dt}=0. \lb{h5.14}
\end{equation}

Next, we turn to the specifics of the Toda hierarchy.
%---------- lemma
\begin{lemma} \lb{lemma5.1}
Assume Hypothesis \ref{h2.1}. Then,
\begin{align}
\frac{\delta \hat{f}_\ell}{\delta a}&=-\frac{\ell}{a}\hat{g}_{\ell-1}, \quad \ell\in\bbN,  
\lb{h5.16}\\
\frac{\delta \hat{f}_\ell}{\delta b}&= \ell \hat{f}_{\ell-1}, \quad \ell\in\bbN.   \lb{h5.15} 
\end{align}
\end{lemma}
%%%%%%%%%%%%%%%%%%%
\begin{proof} With our assumptions on $(a,b)$ we only know that  $\hat{f}_\ell\in\ell^\infty(\bbZ)$.  We start by deriving \eqref{h5.15}.  To that end we introduce the functional
\begin{equation}
\hatt\calF_{\ell,N}(a,b)= \sum_{n\in\bbZ} \hat{f}_\ell(n)\chi_N(n),  \lb{h5.21}
\end{equation}
where $\chi_N$ is the characteristic function of the set $[-N,N]\cap\bbZ$. Then one finds
\begin{align}
&(d\hatt\calF_{\ell,N}(a,b))_b(v)\no \\
&\quad =\sum_{n\in\bbZ} \bigg( \chi_N(n) \frac{\partial \hat{f}_\ell}{\partial b}+ \chi^{-}_N(n) S^{(-1)}\frac{\partial \hat{f}_\ell}{\partial b^{(-1)}}
+  \chi^{+}_N(n) S^{(+1)}\frac{\partial \hat{f}_\ell}{\partial b^{(+1)}}+\cdots\bigg)v(n) \no \\
&\quad\underset{N\to\infty}{\to}\sum_{n\in\bbZ} \bigg(\frac{\partial \hat{f}_\ell}{\partial b}
+  S^{(-1)}\frac{\partial \hat{f}_\ell}{\partial b^{(-1)}}
+ S^{(+1)}\frac{\partial \hat{f}_\ell}{\partial b^{(+1)}}+\cdots\bigg)v(n) \no \\
&\quad=\sum_{n\in\bbZ} \frac{\delta \hat{f}_\ell}{\delta b}(n)v(n), 
\quad  v\in \ell^1(\bbZ).  \lb{h5.22}
\end{align}
On the other hand, recalling that $(\dott,\dott)_{\ell^2(\bbZ)}$ denotes the usual scalar product in $\ell^2(\bbZ)$, 
\begin{align}
&(d\hatt\calF_{\ell,N}(a,b))_b(v)
=\frac{d}{d\epsilon}\hatt\calF_{\ell,N}(b+\epsilon v)\big|_{\epsilon=0}
=  \sum_{n\in\bbZ} \bigg(\delta_n,\sum_{k=0}^{\ell-1}
L^k v L^{\ell-1-k}\delta_n\bigg)_{\ell^2(\bbZ)}\chi_N(n) \no \\
& \;\; =  \sum_{n\in\bbZ} \bigg(\delta_n,\sum_{k=0}^{\ell-1}L^{\ell-1}
v\delta_n\bigg)_{\ell^2(\bbZ)}\chi_N(n)
+\sum_{n\in\bbZ} \bigg(\delta_n,\sum_{k=0}^{\ell-1}
[L^k v,L^{\ell-1-k}]\delta_n\bigg)_{\ell^2(\bbZ)}\chi_N(n) \no \\
&\;\; \underset{N\to\infty}{\to} \sum_{k=0}^{\ell-1} \sum_{n\in\bbZ} \big(\delta_n,L^{\ell-1}\delta_n\big)_{\ell^2(\bbZ)}v(n)
+  \sum_{k=0}^{\ell-1} \sum_{n\in\bbZ} \big(\delta_n,
[L^k v,L^{\ell-1-k}]\delta_n\big)_{\ell^2(\bbZ)}\no \\
&\;\; =\ell \sum_{n\in\bbZ} \big(\delta_n,L^{\ell-1}\delta_n\big)_{\ell^2(\bbZ)}v(n) 
 =\ell \sum_{n\in\bbZ} \hat f_{\ell-1}(n)  v(n) , \quad 
v\in \ell^1(\bbZ),     
\lb{h5.23}
\end{align}
using \eqref{3.4.37}, and the general result that for bounded operators 
$A,B \in \calB(\calH)$ on a separable, complex Hilbert space $\calH$ with $AB$ and 
$BA$ trace class operators,  their commutator is traceless, that is, $\tr([A,B])=0$ (cf.\ 
\cite{De78}, \cite[Corollary 3.8]{Si05}). 
Combining the two expressions, one concludes that \eqref{h5.15} holds.

Next we turn to the proof of \eqref{h5.16}. To this end one first observes that, as before, 
\begin{align}
(d\hatt\calF_{\ell,N}(a,b))_a(v) &
\underset{N\to\infty}{\to}  \sum_{n\in\bbZ} \bigg(\frac{\partial \hat{f}_\ell}{\partial a}
+  S^{(-1)}\frac{\partial \hat{f}_\ell}{\partial a^{(-1)}}
+ S^{(+1)}\frac{\partial \hat{f}_\ell}{\partial a^{(+1)}}+\cdots\bigg)v(n) \no \\
& \quad \;\; =\sum_{n\in\bbZ} \frac{\delta \hat{f}_\ell}{\delta a}(n)v(n), 
\quad  v\in \ell^1(\bbZ).  \lb{h5.22a}
\end{align}
Furthermore,  one computes 
\begin{align}
% \sum_{n\in\bbZ} \frac{\delta  \hat{f}_\ell}{\delta a}(n) v(n)
(d\hatt\calF_{\ell,N}(a,b))_a(v)&= \frac{d}{d\epsilon}\hatt\calF_{\ell,N}(a+\epsilon v)\big|_{\epsilon=0} \no \\
&=\frac{d}{d\epsilon}\sum_{n\in\bbZ} 
\big(\delta_n,\big((a+\epsilon v) S^+ + (a+\epsilon v)^- S^- + b\big)^\ell 
\delta_n\big)_{\ell^2(\bbZ)}\chi_N(n)\Big|_{\epsilon=0} \no \\
&=\sum_{n\in\bbZ}\bigg(\delta_n, 
\sum_{k=0}^{\ell-1}\big(L^k(v S^+ + v^- S^- )L^{\ell-1-k}\big)\delta_n\bigg)_{\ell^2(\bbZ)}\chi_N(n) \no \\
&= \sum_{k=0}^{\ell-1} \sum_{n\in\bbZ} \big(\delta_n,(v S^+ + v^- S^- )L^{\ell-1}\delta_n\big)_{\ell^2(\bbZ)}v(n)\chi_N(n)\no \\
&\quad +  \sum_{k=0}^{\ell-1}\sum_{n\in\bbZ}
\big(\delta_n,[L^k (v S^+ + v^- S^- ),L^{\ell-1-k}]\delta_n\big)\chi_N(n)\no \\
&\underset{N\to\infty}{\to} \sum_{k=0}^{\ell-1} \sum_{n\in\bbZ} \big(\delta_n,(v S^+ + v^- S^- )L^{\ell-1}\delta_n\big)_{\ell^2(\bbZ)}v(n)\no \\
&\qquad +  \sum_{k=0}^{\ell-1}\tr\big([L^k (v S^+ + v^- S^- ),L^{\ell-1-k}]\big)\no \\
&=\ell \sum_{n\in\bbZ}\Big( \big(\delta_n,v S^+L^{\ell-1}\delta_n\big)_{\ell^2(\bbZ)}
+\big(\delta_n, v^- S^- L^{\ell-1}\delta_n\big)_{\ell^2(\bbZ)}\Big)\no \\
&=\ell \sum_{n\in\bbZ}\Big(v(n)\big( S^- \delta_n, L^{\ell-1}\delta_n\big)_{\ell^2(\bbZ)} 
+ v(n-1)\big(S^+\delta_n,   L^{\ell-1}\delta_n\big)_{\ell^2(\bbZ)} \Big)\no \\
&=\ell \sum_{n\in\bbZ}\Big(v(n)\big( \delta_{n+1}, L^{\ell-1}\delta_n\big)_{\ell^2(\bbZ)} 
+ v(n-1)\big(\delta_{n-1},   L^{\ell-1}\delta_n\big)_{\ell^2(\bbZ)} \Big)\no \\
&=\ell \sum_{n\in\bbZ}\Big(v(n)\big(\delta_{n+1}, L^{\ell-1}\delta_n\big)_{\ell^2(\bbZ)} 
+ v(n)\big(\delta_{n},   L^{\ell-1}\delta_{n+1}\big)_{\ell^2(\bbZ)} \Big)\no \\
&= 2\ell\sum_{n\in\bbZ} \big(\delta_{n+1}, L^{\ell-1}\delta_n\big)_{\ell^2(\bbZ)} v(n)\no \\
&=-\ell \sum_{n\in\bbZ}\frac{1}{a(n)}  \hat{g}_{\ell-1}(n)v(n), \quad 
v\in \ell^1(\bbZ).   \lb{h5.20}
\end{align}
Thus \eqref{h5.16} is established.
\end{proof}
%------------ end lemma

For the remainder of this section we now introduce the following assumption. 
%\marginlabel{\bf{Alternative proof if $\ell$ even.}}

%%%%%%%%%%%%%%%%%%%%%%%%%%%%%%%%%%%%% 
\begin{hypothesis} \lb{hh9.9}
In addition to Hypothesis \ref{h2.1} suppose that
\begin{equation}
\big\{a(n)^2-\tfrac{1}{4}\big\}_{n\in\bbZ}, \{b(n)\}_{n\in\bbZ} \in \ell^1(\bbZ).  \lb{h5.24}
\end{equation}
\end{hypothesis}
%%%%%%%%%%%%%%%%%%%%%%%%%%%%%%%%%%%%% 

We fix $\ell \in \bbN$ and want to show that a suitably renormalized functional 
$\hatt\calF_{\ell}$ is well-defined.  We define it by subtracting the limit of each  
$\hat{f}_\ell(n)$ as $n\to\infty$.
Each $\hat{f}_\ell$ is a polynomial in $a,b$ and certain shifts thereof.  Terms with $a$ and shifts enter homogeneously and only in even powers.  Assume that $\hat{f}_\ell(n)\to\lambda_\ell$ as $\abs{n}\to\infty$.  For $\ell$ odd, we know that $\lambda_\ell=0$ because each each term contains at least one 
$b$ or certain shifts thereof. Then 
$\{\hat{f}_\ell(n)-\lambda_\ell\}_{n\in\bbZ}\in \ell^1(\bbZ)$.  We can see this as follows. Each 
$\hat{f}_\ell$ is a finite sum of terms containing $a$ and $b$ with shifts. Terms with $b$ are already summable.  Only terms exclusively with $a$ have a nonzero limit and hence are nonsummable. Their general form will be 
$\alpha_1(n)\cdots \alpha_k(n)$ with $\alpha_j(n)=(a^{(m_j)}(n))^{2p_j}\to\tilde\lambda_j$ as $\abs{n}\to\infty$, where $(m_j)$ denotes the shifts. Then
\begin{align}
\alpha_1(n)\cdots \alpha_k(n)-\tilde\lambda_1\cdots\tilde\lambda_k
&= (\alpha_1(n)-\tilde\lambda_1)\alpha_2(n)\cdots \alpha_k(n)\no\\
&\quad+
\tilde\lambda_1(\alpha_2(n)-  \tilde\lambda_2) \alpha_2(n)\cdots \alpha_k(n) \no\\
&\quad +\cdots+\tilde\lambda_1\cdots\tilde\lambda_{k-1} (\alpha_k(n)-\tilde\lambda_k).      
\end{align}
Thus we see that 
\begin{equation}
\{\alpha_1(n)\cdots \alpha_k(n)-\tilde\lambda_1\cdots\tilde\lambda_k \}_{n\in\bbZ}
\in\ell^1(\bbZ),
\end{equation}
and hence 
\begin{equation}
\{\hat{f}_\ell(n)-\lambda_\ell\}_{n\in\bbZ}\in\ell^1(\bbZ).
\end{equation}
Thus, the functional 
\begin{equation}
\hatt\calF_{\ell}(a,b)=\sum_{n\in\bbZ} \big(\hat{f}_\ell(n)-\lambda_\ell\big)
\end{equation}
is well-defined.  

In the following it is convenient to introduce the abbreviation 
$\bbN_{-1}=\bbN\cup\{-1,0\}$.

%------------ theorem
\begin{theorem} \lb{thm:5.4}
Assume Hypothesis \ref{hh9.9} and let $p\in\bbN_{-1}$. In addition, define 
\begin{equation}
\widehat I_{p}= \frac{1}{p+2}\sum_{n\in\bbZ} \big(\hat{f}_{p+2}(n)-\lambda_{p+2}\big),  
\lb{h5.24a}
\end{equation}
where 
\begin{equation}
\lambda_{p+2}=\lim_{\abs{n}\to\infty}\hat f_{p+2}(n). 
\end{equation}
Then,
\begin{equation}
(\nabla \widehat I_p)_a=  \frac{1}{p+2}\frac{\delta \hat{f}_{p+2}}{\delta a}
=-\frac{1}{a} \hat{g}_{p+1}, \quad 
(\nabla \widehat I_p)_b=  \frac{1}{p+2}\frac{\delta \hat{f}_{p+2}}{\delta b}= \hat{f}_{p+1}.
\end{equation}
Moreover, the homogeneous $p$th Toda equations then take on the form 
\begin{align}
\htl_p(a,b)= \begin{pmatrix}a_{t_p}\\  b_{t_p}\end{pmatrix}
 - D \begin{pmatrix}(\nabla \widehat I_p)_a\\[1mm]   (\nabla  \widehat I_p)_b\end{pmatrix} =0,   \quad p\in\bbN_{-1}.    \lb{h5.38}
\end{align}
\end{theorem}
\noindent 
%%%%%%%%%%%%%%%%%
\begin{proof} 
This follows directly from Lemma  \ref{lemma5.1}  and \eqref{3.2.13}.
\end{proof}
%------------ end theorem
%%%%%%%%%%%%%%%%%

%%%%%%%%%%%%%%%%%
%------------ begin ---------------
\begin{remark}
In \eqref{h5.38} we also introduced the trivial linear flow $\htl_{-1}$ given by
\begin{equation}
\htl_{-1}(a,b)= \begin{pmatrix}a_{t_{-1}}\\  b_{t_{-1}}\end{pmatrix}
 - D \begin{pmatrix}(\nabla \widehat I_{-1})_a\\[1mm]   (\nabla  \widehat I_{-1})_b\end{pmatrix} 
 = \begin{pmatrix}a_{t_{-1}}\\  b_{t_{-1}}\end{pmatrix} =0.
\end{equation}
\end{remark}
%------------------- end remark
%%%%%%%%%%%%%%%%%

Next, we consider the general case, where
\begin{equation}
\tl_p(a,b) = \begin{pmatrix}  a_{t_p}-a(f_{p+1}^+-f_{p+1}) \\[1mm]
b_{t_p}+g_{p+1}-g_{p+1}^- \end{pmatrix} =0, \quad p\in\bbN_{-1},  \lb{3.2.13a}
\end{equation}
and
\begin{equation}
f_p=\sum_{k=0}^{p}c_{p-k}\hat{f}_{k},\quad
g_p=\sum_{k=1}^{p}c_{p-k}\hat{g}_{k}-c_{p+1}, \quad p\in\bbN_0. \lb{1.2.7a}
\end{equation}
Introducing  
\begin{equation}
I_{-1} = \widehat I_{-1}, \quad 
I_p= \sum_{k=0}^p c_{p-k} \widehat I_k,   \quad p\in\bbN_0,  \lb{h5.44}
\end{equation}
one then obtains
\begin{equation}
\tl_p(a,b) = \begin{pmatrix}a_{t_p}\\  b_{t_p}\end{pmatrix}
 - D \begin{pmatrix}(\nabla I_p)_a\\[1mm]  (\nabla I_p)_b\end{pmatrix} =0, 
 \quad p\in\bbN_{-1}.   \lb{h5.45}
\end{equation}

%%%%%%%%%%%%%%%%%
%------------ theorem
\begin{theorem} \lb{thm5.5}
Assume Hypothesis \ref{hh9.9} and suppose that $a, b$ satisfy $\tl_p(a,b)=0$ for some 
$p\in\bbN_{-1}$. Then, 
\begin{equation}
\frac{d I_r}{dt_p} =0, \quad r\in\bbN_{-1}.   \lb{h9.60}
\end{equation}
\end{theorem}
%%%%%%%%%%%
\begin{proof} From \eqref{h3.2} and \eqref{h2.15} (cf.\ Remark \ref{rem4.2}) one obtains 
\begin{equation}
\frac{d\hat f_{r+2}}{dt_p}=(S^+ -I)\beta_{r+2}, \quad r\in\bbN_{-1}, 
\end{equation}
for some $\beta_{r+2}$, $r\in\bbN_{-1}$, which are polynomials in $a$ and $b$ and certain shifts thereof. Using definition \eqref{h5.24a} of $\hatt I_r$,  the result \eqref{h9.60} follows in the homogeneous case and then by linearity in the general case. 
\end{proof}
%------------ end theorem
%%%%%%%%%%%%%%%%%

%%%%%%%%%%%%%%%%%
%------------ theorem
\begin{theorem} \lb{ht5.6}
Assume Hypothesis \ref{hh9.9} and let $p,r\in\bbN_{-1}$. Then, 
\begin{equation}
\{I_p, I_r\}=0,   \lb{h9.61}
\end{equation}
that is, $I_p$ and $I_r$ are in involution for all $p,r \in \bbN_{-1}$. 
\end{theorem}
%%%%%%%%%%%%%%%%%
\begin{proof}  By Remark \ref{r9.7}\,$(i)$, there exists $T_0>0$ such that the initial value problem 
\begin{equation}
\tl_{p}(a,b)=0, \quad (a,b)\big|_{t=0}=\big(a^{(0)},b^{(0)}\big),
\end{equation}
where $a^{(0)}, b^{(0)}$ satisfy Hypothesis \ref{hh9.9}, has a unique and continuous solution $a(t), b(t)$ satisfying Hypothesis \ref{hh9.9} for each $t\in[0,T)$. For this solution we know that
\begin{equation}
\frac{d}{dt} I_p(t)=\{I_r(t), I_p(t)\}=0.
\end{equation}
Next, let $t\downarrow 0$. Then
\begin{equation}
0=\{I_r(t), I_p(t)\}\underset{t\downarrow 0}{\to}\{I_r(0), I_p(0)\}
=\{I_r, I_p\}\big|_{(a,b)=(a^{(0)},b^{(0)})}. 
\end{equation}
Since $a^{(0)}, b^{(0)}$ are arbitrary coefficients satisfying Hypothesis \ref{hh9.9} one concludes \eqref{h9.61}.
\end{proof}
%------------ end theorem
%%%%%%%%%%%%%%%%%%%

There is also a second Hamiltonian structure for the Toda hierarchy.  Rewriting the linear recursion \eqref{1.2.4a}--\eqref{1.2.4c} in the form
\begin{align}
f_{\ell+1}^+ - f_{\ell+1}&=-\frac12(g_{\ell}^+ - g_{\ell}^-)+(b^+ f_{\ell}^+ - bf_{\ell}),\no  \\
g_{\ell+1} - g_{\ell+1}^-&=-2(a^2 f_{\ell}^+ - (a^-)^2f_{\ell}^-)+b(g_{\ell} - g_{\ell}^-),  
\quad \ell\in\bbN_0,   \lb{h5.51}
\end{align}
one finds
\begin{equation}
\begin{pmatrix} \frac12(f_{n+1}^+ - f_{n+1})\\ g_{n+1}^-  - g_{n+1} \end{pmatrix}
=\widetilde D \begin{pmatrix} -\frac1a g_n \\ f_n\end{pmatrix},  \lb{h5.51a}
\end{equation}
where we abbreviated
\begin{equation}
\widetilde D=\begin{pmatrix}\frac12 a(S^+ - S^-)a & a(S^+ -I)b \\[1mm]
b(I-S^-)a & 2(S^+ - S^-)a^2  \end{pmatrix}.   \lb{h5.52}
\end{equation}
One can introduce a second Poisson bracket $\{\{\dott,\dott\}\}$, defined by
\begin{equation}
\{\{ \calF_1,\calF_2\}\}  = \sum_{n\in\bbZ}
\begin{pmatrix}\frac{\delta F_1}{\delta a}(n)\\[1.5mm] \frac{\delta F_1}{\delta b}(n)\end{pmatrix} \cdot 
\widetilde D
\begin{pmatrix}\frac{\delta F_2}{\delta a}(n)\\[1.5mm] \frac{\delta F_2}{\delta b}(n)\end{pmatrix}. 
\lb{h5.53}
\end{equation}

This second Poisson bracket is also skew-symmetric and satisfies the Jacobi identity and the Leibniz rule. As in Theorem \ref{ht5.6}, one verifies that all $I_p$, $p\in\bbN_{-1}$, 
 are in involution also with respect to this second Poisson bracket \eqref{h5.53}, that is, 
\begin{equation}
\{\{I_r,I_p\}\}=0, \quad p,r \in \bbN_{-1}. 
\end{equation}
Combining \eqref{h5.24a}, \eqref{h5.44}, \eqref{h5.45}, and \eqref{h5.51}, \eqref{h5.52} permits one to write the $p$th Toda equation in another Hamiltonian form
\begin{align}
\begin{split}
\tl_p(a,b)&= \begin{pmatrix}a_{t_p}\\  b_{t_p}\end{pmatrix} -\widetilde D 
\begin{pmatrix}(\nabla I_{p-1})_a\\[1mm]  (\nabla I_{p-1})_b\end{pmatrix} =0, 
\quad p \in \bbN_{0}.   \lb{h5.54}
\end{split}
\end{align}
%\fbox{\bf What about $\tl_{-1}(a,b)$ ??????}

%%%%%%%%%%%%%%%%%%%%%%%%%%%%%%%%%%%%
%%%%%%%%%%%%%%%%%%%%%%%%%%%%%%%%%%%%
\section{A sketch of the almost periodic case} \lb{s6}
%%%%%%%%%%%%%%%%%%%%%%%%%%%%%%%%%%%%
%%%%%%%%%%%%%%%%%%%%%%%%%%%%%%%%%%%%

Following Johnson and Moser \cite{JM82} in the context of one-dimensional almost periodic 
Schr\"odinger operators and Carmona and Kotani \cite{CK87} in the corresponding discrete  case (see also the treatment in Carmona and Lacroix 
\cite[Sects.\ VII.1, VII.2]{CL90}), we now very briefly sketch an extension of the Hamiltonian formalism to the case of almost periodic coefficients $a$ and $b$.

To set the stage, if $c$ denotes an almost periodic sequence on $\bbZ$, we recall 
that its ergodic mean $\langle c\rangle$ is given by
\begin{equation}
\langle c\rangle = \lim_{N\uparrow\infty} \f{1}{2N+1} \sum_{n=-N}^N c_n. 
\end{equation}
We assume that $u$ has the frequency module $\calM(u)$ and given a density $F$ as in \eqref{h5.3A}, equation \eqref{h5.1A} then becomes
\begin{align}
\begin{split}
\calF(u) &=\lim_{N\uparrow\infty}\sum_{n=-N}^N 
F\big(n,u(n), u^{(+1)}(n), u^{(-1)}(n), \dots, u^{(k)}(n), u^{(-k)}(n)\big), \\
&= \langle F(\dott,u) \rangle, 
\end{split}
\end{align}
and assuming that the frequency module $\calM(v)$ of $v$ satisfies $\calM(v)\subseteq \calM(u)$, the analog of \eqref{h5.4a} then reads
\begin{align}
(d\calF)_u (v)&=\frac{d}{d\epsilon}\calF(u+\epsilon v)\big|_{\epsilon=0} \no \\ 
&=  \lim_{N\uparrow\infty}\sum_{n=-N}^N 
\bigg(\frac{\partial F(n,u)}{\partial u}+ S^{(-1)}\frac{\partial F(n,u)}{\partial u^{(+1)}} + 
S^{(+1)}\frac{\partial F(n,u)}{\partial u^{(-1)}}\no \\
&\hspace*{2.35cm } +\cdots+S^{(-k)}\frac{\partial F(n,u)}{\partial u^{(k)}} 
+S^{(k)}\frac{\partial F(n,u)}{\partial u^{(-k)}} \bigg)v(n) \no \\
&= \big\langle (\nabla \calF)_u v \big\rangle
= \bigg\langle \f{\delta F}{\delta u} v \bigg\rangle. 
\end{align}

Assuming $a$ and $b$ are almost periodic with frequency module $\calM$, we again  consider functionals as in \eqref{h5.3a},
\begin{align}
& \calF(a,b) = \big\langle F(\dott,a,b, a^{(+1)}, b^{(+1)},a^{(-1)}, b^{(-1)},\dots, a^{(+k)}, b^{(+k)},a^{(-k)}, b^{(-k)}) \big\rangle  \no \\
& \qquad\quad  = \big\langle F(\dott,a,b) \big\rangle,  
\end{align}
and in analogy to \eqref{h5.8}, the Poisson brackets of two functionals $\calF_1, \calF_2$ are then given by
\begin{equation}
\{\calF_1,\calF_2\}= \Bigg\langle 
\begin{pmatrix}\frac{\delta F_1}{\delta a}(n)\\[1.5mm] \frac{\delta F_1}{\delta b}(n)\end{pmatrix} \cdot 
D\begin{pmatrix}\frac{\delta F_2}{\delta a}(n)\\[1.5mm] \frac{\delta F_2}{\delta b}(n)\end{pmatrix} \Bigg\rangle.
\end{equation}
Again one verifies that both the Jacobi identity as well as the Leibniz rule hold in this case. Moreover, as in \eqref{h5.11} and \eqref{h5.12}, if $\calF$ is a smooth functional and $(a,b)$ develops according to a Hamiltonian flow  with Hamiltonian $\calH$, that is,
\begin{equation}
\begin{pmatrix} a\\ b\end{pmatrix}_t=(\nabla_s\calH)_{a,b}=D(\nabla\calH)_{a,b} 
=D\begin{pmatrix} \frac{\delta H}{\delta a}\\[1mm] \frac{\delta H}{\delta b}\end{pmatrix},
\end{equation}
then
\begin{equation}
\frac{d\calF}{dt}=\frac{d}{dt}\langle F(n,a,b) \rangle  = \{\calF,\calH\}.  
\end{equation}

Next, assuming in addition that $1/a\in\ell^\infty(\bbN)$ and that $\langle 1/a\rangle $ stays finite for $t$ varying in some open time interval $\bbI\subseteq\bbR$, we now introduce the fundamental function $w$ by
\begin{equation}
w(z) = \langle \phi_+(z,\dott)\rangle  \lb{h6.8}
\end{equation}
for $|z|$ sufficiently large. Since 
\begin{equation}
w'(z)= - \langle g(z,\dott) \rangle, \quad z \in \bbC \setminus \spec(H),
\end{equation}
$w$ in \eqref{h6.8} extends analytically to $z\in \bbC \setminus \spec(H)$. We note that $w$ is the appropriate discrete analog of the function $\langle m_+(z) \rangle = 
\langle -1/G(z,\dott,\dott)\rangle$ introduced by Johnson and Moser \cite{JM82} in the case of almost periodic Schr\"odinger operators. (Here $m_+$ denotes the right half-line Weyl--Titchmarsh coefficient and $G(z,\dott,\dott)$ the diagonal Green's function of the underlying almost periodic Schr\"odinger operator $H=-d^2/dx^2 +V$ in $L^2(\bbR)$.)

One observes the asymptotic expansion 
\begin{align}
\begin{split}
w(z) = \langle \ln(\phi_+(z,\dott)\rangle &\underset{z\to\infty}{=} \langle \ln(a/z) \rangle 
+ \sum_{j\in\bbN} \langle \rho_{+,j} \rangle z^{-j}  \\
& \underset{z\to\infty}{=} \langle \ln(a/z) \rangle 
+ \sum_{j\in\bbN} j^{-1} \big\langle \hat f_j \big\rangle z^{-j} 
\end{split}
\end{align}
and computes
\begin{align}
\f{\delta w(z)}{\delta a} (n) & = \f{1}{a(n)} - 2 G(z,n,n+1), 
\quad z \in \bbC \setminus \spec(H), \; n\in\bbZ,  \\
\f{\delta w(z)}{\delta b} (n) & = -g(z,n), \quad z \in \bbC \setminus \spec(H), \; n\in\bbZ. 
\end{align}

Introducing
\begin{equation}
\widehat I_{-1} = I_{-1} = \big\langle \hat f_1\big\rangle, \quad 
\hatt I_p =\f{1}{p+2} \big\langle \hat f_{p+2}\big\rangle, \quad 
I_p = \sum_{k=0}^p c_{p-k} \hatt I_k, \quad p\in\bbN_0, 
\end{equation}
the Toda equations again take on the form 
\begin{equation}
\tl_p(a,b) = \begin{pmatrix}a_{t_p}\\  b_{t_p}\end{pmatrix}
 - D \begin{pmatrix}(\nabla I_p)_a\\[1mm]  (\nabla I_p)_b\end{pmatrix} =0,   
  \quad p \in \bbN_{-1}, 
\end{equation}

Finally, one can show that $w(z_1)$ and $w(z_2)$ are in involution for arbitrary 
$z_1, z_2 \in \bbC \setminus \spec(H)$, and hence obtains
\begin{align}
& \{w(z_1), w(z_2)\} =0,  \quad z_1, z_2 \in \bbC \setminus \spec(H),   \\
& \{w(z), I_p\} =0,  \quad \{I_p, I_r\} =0, \quad z \in \bbC \setminus \spec(H), \; 
p,r \in\bbN_{-1}. 
\end{align}

Naturally, these considerations apply to the special periodic case in which $\langle c\rangle$ for a periodic sequence $c$ on $\bbZ$ is to be interpreted as the periodic mean value.

%%%%%%%%%%%%%%%%%%%%%%%%%%%%%%%%%%%%%%%
\bigskip
\noindent {\bf Acknowledgments.}
Fritz Gesztesy gratefully acknowledges the extraordinary
hospitality of the Department of Mathematical Sciences of
the Norwegian University of Science and Technology, Trondheim, during 
two-month stays in the summers of 2005 and 2006, where parts of this paper
were written. He is also indebted to the Office of Research of the University of
Missouri--Columbia, for granting a research leave for the
academic year 2005/06. Moreover, we gratefully acknowledge the hospitality of the Mittag-Leffler Institute, Sweden, creating a great working environment for research, during the fall of 2005.
%%%%%%%%%%%%%%%%%%%%%%%%%%%%%%%%%%%%%%%%

%%%%%%%%%%%%%%%%%%%%%%%%%%%%%%%%%%%%%%%%%-------------------
% when bibtexing please active the next two lines
%\bibliographystyle{plainMOD}
%\bibliography{solitonref}
%------- end bibtex
% after bibtex, please paste file CH.bbl into this file below here
% and comment away the (first) two lines above.
%%%%%%%%%%%%%%%%%%%%%%%%%%%%%%%%%%%%%%%%%
%%%%%%%%%%%%%%%%%%%%%%%%%%%%%%%%%%%%%%%%%  

\end{document}